\documentstyle[11pt,epsfig]{article}
\newcommand{\fbar}{\not{\!f}}

\newcommand{\Qbar}{\not{\!Q}}
\newcommand{\Kbar}{\not{\!K}}

\newcommand{\Hslash}{\not{\!H}}
\newcommand{\Pbar}{\not{\!P}}

\newcommand{\be}{\begin{equation}}
\newcommand{\ee}{\end{equation}}
\newcommand{\ba}{\begin{eqnarray}}
\newcommand{\ea}{\end{eqnarray}}

\newcommand{\nsigma}{\mbox{\boldmath $\sigma$}}
\newcommand{\ngamma}{\mbox{\boldmath $\gamma$}}

\newcommand{\nh}{{\bf      h}}

\newcommand{\nk}{{\bf      k}}
\newcommand{\np}{{\bf      p}}       
\newcommand{\nq}{{\bf      q}}

\newcommand{\columnmatrix}[2]{\left[
                               \begin{array}{cc}
                                  \displaystyle #1 \\[1ex]
                                  \displaystyle #2
                               \end{array}
                            \right]}
\begin{document}
\begin{titlepage}
\mbox{} 
\vspace*{2.5\fill} 
{\Large\bf 
\begin{center}
%
Consistent one-pion exchange currents 
in electron scattering from a
relativistic Fermi gas
%
\end{center}
} 
\vspace{1\fill} 
\begin{center}
{\large 
J.E. Amaro$    ^{1}$, 
M.B. Barbaro$  ^{2}$, 
J.A. Caballero$^{3,4}$, 
T.W. Donnelly$ ^{5}$ and 
A. Molinari$   ^{2}$
}
\end{center}
\begin{small}
\begin{center}
$^{1}${\sl 
Departamento de F\'\i sica Moderna,
Universidad de Granada, 
E-18071 Granada, SPAIN 
}\\[2mm]
$^{2}${\sl 
Dipartimento di Fisica Teorica,
Universit\`a di Torino and
INFN, Sezione di Torino \\
Via P. Giuria 1, 10125 Torino, ITALY 
}\\[2mm]
$^{3}${\sl 
Departamento de F\'\i sica At\'omica, Molecular y Nuclear \\ 
Universidad de Sevilla, Apdo. 1065, E-41080 Sevilla, SPAIN 
}\\[2mm]
$^{4}${\sl 
Instituto de Estructura de la Materia, CSIC 
Serrano 123, E-28006 Madrid, SPAIN 
}\\[2mm]
$^{5}${\sl 
Center for Theoretical Physics, Laboratory for Nuclear Science 
and Department of Physics\\
Massachusetts Institute of Technology,
Cambridge, MA 02139, USA 
}
\end{center}
\end{small}

\kern 1. cm \hrule \kern 3mm 

\begin{small}
\noindent
{\bf Abstract} 
\vspace{3mm} 

A set of fully relativistic one-pion-exchange electromagnetic
operators is developed for use in one-particle emission reactions
induced by electrons. To preserve the gauge invariance of the theory
additional pionic correlation operators are required beyond the usual
meson-exchange current operators. Of these, in the present work
emphasis is placed on the self-energy current which is infinite and
needs to be renormalized. The renormalized current is expanded to
first order to obtain a genuine one-pion-exchange contribution to the
inclusive responses.

\kern 2mm 

\noindent
{\em PACS:}\  25.30.Rw, 14.20.Gk, 24.10.Jv, 24.30.Gd, 13.40.Gp  
\noindent
{\em Keywords:}\ Nuclear reactions; Inclusive electron scattering;
Pionic correlations. Meson-exchange currents. Relativistic Fermi Gas.

\end{small}
\kern 2mm \hrule \kern 1cm
\noindent MIT/CTP\#3167 
\end{titlepage}

\section{Introduction}

In a recent paper~\cite{Ama01} we investigated the role played by
pions in inclusive electron scattering from nuclei within the context
of one-particle one-hole (1p-1h) excitations. There we extended our
previous work~\cite{Ama98,Ama99} in which we had pursued a systematic
study of relativistic effects in the nuclear electromagnetic responses
for various kinematical regions incorporating both meson-exchange and
isobar currents. The resulting two-body current is a consistent
first-order operator, since it contains all Feynman diagrams built
from nucleons and pions with one exchanged pion and with a photon
attached to all possible lines; importantly, the result is gauge
invariant, as explicitly proven in \cite{Ama01}.

In addition to the usual contact and pion-in-flight meson-exchange
current (MEC) operators, this fully relativistic operator includes the
so-called correlation operator.  Usually this is not included in model
calculations because it gives rise to contributions which are, at
least in part, already accounted for in the starting nuclear wave
functions.  However, our model is based on an uncorrelated
relativistic Fermi gas where the ground state is a Slater determinant
constructed with (Dirac) plane waves. Within a perturbative approach
we are free to consider the one-pion correlation contribution to the
responses as arising either explicitly in the wave functions or from a
current operator acting on unperturbed states --- our approach is the
latter.  When the whole perturbative expansion is summed up one must
recover the results obtained starting with an exact ``correlated''
wave function.

In this paper we enter in more depth into analysis of the ideas
in~\cite{Ama01}, focusing on the impact of pions on the nuclear
electromagnetic response in the 1p-1h channel, since this gives the main
contribution for quasielastic
conditions~\cite{Alb90}--\cite{Ama94}. As in the MEC case, the
two-body correlation current also contributes in this channel and thus
is required to obtain a consistent, gauge invariant, one-pion-exchange
current.  When acting on the ground state with the two-body operator,
the 1p-1h matrix element associated with the two-body currents is
obtained by integrating one of the two particle states over the Fermi
sea. In so-doing two kinds of contributions are obtained. The first
one, sometimes referred to as a vertex correction~\cite{Blu89}, arises
from the pionic correlations between the particle and hole; the second
relates to the Fock self-energy (s.e.)~\cite{Blu89,Hor90} and dresses
the particle and hole lines. This second contribution diverges, since
it corresponds to a s.e. insertion on an external line, which,
according to field theory~\cite{Bjo65,Pes95}, should not be included
in a perturbative expansion. Instead one should apply a
renormalization procedure to dress the external lines by summing the
entire perturbative series of self-energy insertions. In the nuclear
case this procedure leads to the relativistic Hartree-Fock (HF)
approach.

In other relativistic calculations~\cite{Blu89,Hor90}
this contribution has been treated by introducing from the start
a Hartree-Fock propagator in
the medium, which accounts for the s.e. diagrams.
In-medium form factors for the 1p-1h current are also defined, 
however neglecting any momentum
dependence in the self-energy and effective mass. In \cite{Hor90} 
the self-consistent Hartree mean field was inserted into
the single-particle propagator, automatically including 
the Pauli blocking of $N\overline{N}$ pairs, whose contribution was thus
included in the RPA responses.
A similar treatment including a semi-phenomenological nucleon 
self-energy  in the medium at the non-relativistic level was 
implemented in ref. \cite{Gil97}.

In~\cite{Ama01} the difficulty posed by the s.e. insertion in first
order was avoided by computing the self-energy response as the
imaginary part of the corresponding polarization propagator with
one-pion-exchange s.e.  insertions on the particle and hole lines. In
this way a finite result is already obtained in first order (one
pionic line) without having to resort to the HF approach.  The
question then arises whether it is possible to obtain the same result
for the self-energy response function starting with {\em finite
well-defined} matrix elements of the current operator.

In this paper our aim is to answer this question by constructing a
renormalized self-energy current corresponding to one-pion-exchange.
This current acts over free Dirac spinors and leads to the same
response functions as those obtained by taking the imaginary part of
the polarization propagator computed to first order.  It should be
clear that here the concept of renormalization has a many-body
significance, namely it amounts to a relativistic HF approximation and
ignores (see~\cite{Ama01}) the additional vacuum renormalization due
to the change of the negative-energy sea induced by the nuclear
medium~\cite{Hor90}.

The new current is obtained by renormalizing spinors and energies and
expanding the resulting in-medium one-body current to first order in
the square of the pion-nucleon coupling constant. The renormalized
quantities should be obtained by solving a set of self-consistent
relativistic HF equations numerically. However, to first order in the
expansion, it is possible to write down analytic expressions for the
solutions and the corresponding corrections to the bare single-nucleon
current operator can be expressed in terms of a simple electromagnetic
operator. This operator accounts for two main effects introduced by
the interaction of the nucleon with the medium: the first is the
enhancement of the lower-components of Dirac spinors; the second is a
global renormalization of the spinors in the nuclear medium.  These
effects are genuine relativistic corrections that are absent within an
essentially
non-relativistic approach~\cite{Alb90}. Actually a third type of
renormalization effect arises from the modification of the
energy-momentum relation for a nucleon, treated in first order of the
square of the pion-nucleon coupling constant (in other approaches this
effect is embedded in a constant effective mass~\cite{Blu89}).

Using the renormalized s.e.  current operator together with the MEC
and the vertex exchange operator we prove that the full current is
gauge invariant if account is taken of the change in energy arising
from the HF renormalization to first order. In this way the results
for the inclusive response functions agree completely with the ones
obtained in \cite{Ama01} with the polarization propagator technique.

The present paper is organized as follows: In Section~2 we briefly
revisit the full set of 1p-1h current operators with one pion-exchange
line which contribute to the electro-excitation process, paying
special attention to the s.e. contribution. We show the necessity of
re-defining the otherwise infinite self-energy diagrams. In Section~3
we perform the renormalization program, thus obtaining the
relativistic HF equations.  In Section~4 we expand the renormalized
spinors and energies to first order in the pion coupling constant
squared obtaining a new self-energy current. We also prove the gauge
invariance of the whole set of current operators to first order.  In
Section~5 we compute the interference response of the s.e. current
with the one-body current and prove its identity with the one derived
in the polarization propagator framework. We present
numerical results for the response
functions focusing on the contributions provided by the different
pieces of the self-energy current.  Finally, in Section~6 we summarize
our results and draw our conclusions.

\section{MEC and Correlation currents}

The linked, two-body Feynman diagrams that contribute to 
electron scattering with one pion-exchange are shown in Fig.~1. The first
three correspond to the usual MEC: diagrams (a), (b)
refer to the contact or seagull current, diagram (c) to the
pion-in-flight current. The four diagrams (d)--(g) represent the
so-called correlation current and are usually not treated as genuine
MEC, but as correlation corrections to the nuclear wave function.
However, again we note that our approach puts all correlation 
effects in the current operator and uses an uncorrelated wave function for
the initial and final nuclear states.

In this work we use Bjorken \& Drell conventions~\cite{Bjo65} and 
a pseudo-vector $\pi N N$ coupling, namely
\begin{equation}
{\cal H}_{\pi NN} = 
\frac{f}{m_\pi}\overline{\psi}\gamma_5\gamma^{\mu}
(\partial_{\mu}\phi_a)\tau_a \psi \ ,
\end{equation}
where $\psi$ is the nucleon field, $\phi_a$ is the isovector pion
field, $f$ represents the $\pi NN$ coupling constant and $m_\pi$ is 
the pion mass. The electromagnetic currents
corresponding to diagrams (a)--(g) are obtained by computing the
$S$-matrix element 
\begin{equation}
S_{fi} = S_{fi}(P'_1,P'_2,P_1,P_2)-S_{fi}(P'_1,P'_2,P_2,P_1)
\end{equation}
for the absorption of a virtual photon
by a system of two nucleons, namely for the process
\be
\gamma + N_1+ N_2 \rightarrow N_1' + N_2'\ ,
\ee
with $P_1$, $P_2$ ($P'_1$, $P'_2$) being the initial (final)
four-momenta of the two nucleons involved. The electromagnetic current 
is then defined according to
\begin{equation}\label{S-matrix}
S_{fi}(P'_1,P'_2,P_1,P_2)=
-ie A_{\mu}(Q) (2\pi)^4 \delta(P'_1+P'_2-P_1-P_2-Q)
j^{\mu}(P'_1,P'_2,P_1,P_2)\ ,
\end{equation}
where $A_{\mu}(Q)$ is related to the matrix element of the
electromagnetic field between the incident photon with momentum $Q$
and the vacuum state, namely
\begin{equation}
\langle 0| A_{\mu}(X)|\gamma(Q)\rangle=
A_{\mu}(Q) e^{-iQ\cdot X}\ .
\end{equation}
In the above $j^{\mu}(P'_1,P'_2,P_1,P_2)$ is related to the matrix
element of the current as follows
\begin{equation}
\langle P'_1 P'_2 | j^{\mu}(Q)| P_1 P_2 \rangle = 
(2\pi)^3 \delta^3(\np'_1+\np'_2-\nq-\np_1-\np_2)
j^{\mu}(P'_1, P'_2,P_1,P_2)\ .
\end{equation}

\begin{figure}[tb]
\begin{center}
\leavevmode
\epsfbox[200 450 400 720]{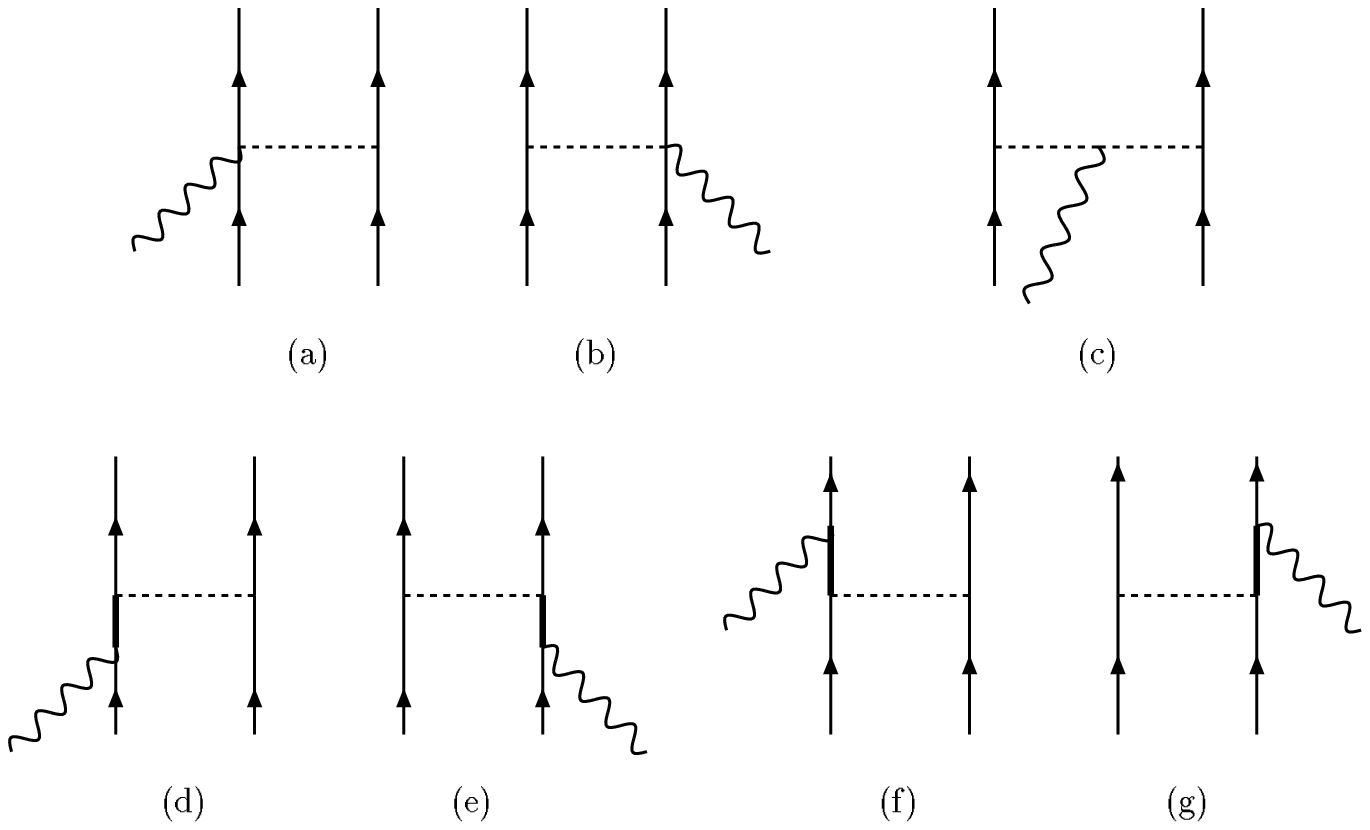}
\end{center}
\caption{Feynman diagrams contributing to the two-body current 
with one pion-exchange. The wide line in the correlation diagrams
(d)--(g) means a fully relativistic Dirac propagator for the nucleon.}
\end{figure}

The general relativistic expressions for the seagull (diagrams a,b),
pion-in-flight (c) and correlation (d-g) currents 
($j^\mu_s$, $j^\mu_p$, $j^\mu_{cor}$) are reported in~\cite{Ama01}.
There we have also proven that when the seagull and pion-in
flight currents are multiplied by the same electromagnetic form factor
$F_1^V$, gauge invariance is fullfilled, {\it i.e.}, 
$Q_{\mu}(j^{\mu}_{s}+j^{\mu}_{p}+j^{\mu}_{cor})=0$.

In this work we deal with the case of one-particle emission 
induced by the two-body currents mentioned above. The matrix element
of a two-body operator between the Fermi gas ground state and a 1p-1h
excited state has the form
\begin{eqnarray}
\langle ph^{-1}|j^{\mu}(Q)|F\rangle 
&\equiv&
(2\pi)^3 \delta^3(\nq+\nh-\np)\frac{m}{V\sqrt{E_\np E_\nh}}j^{\mu}(\np,\nh)
\nonumber\\
&=&
 \sum_{k<F} 
\left[ \langle pk |j^{\mu}(Q)|hk\rangle
      -\langle pk |j^{\mu}(Q)|kh\rangle
\right] \ ,
\label{jmuph}
\end{eqnarray}
where the summation runs over all occupied levels in the ground
state, and thus includes a sum over spin and isospin and an integral over
the momentum $\nk$.
Note that the current matrix element
$j^{\mu}(\np,\nh)$  does not contain the factor
$\frac{m}{\sqrt{E_\np E_\nh}}$, which should accordingly be
included into the phase space when computing the response
functions.  

It is well-known (see, e.g., \cite{Ama01,Ama98}) that the direct term
in eq.~(\ref{jmuph}) vanishes for the MEC and pionic correlation
currents upon summation over the occupied states. Only the exchange
term contributes to the p-h matrix elements. The associated many-body
Feynman diagrams are displayed in Fig.~2 and the corresponding
fully relativistic expressions for the seagull,
pion-in-flight and correlation currents are given in~\cite{Ama01}.

\begin{figure}[tb]
\begin{center}
\leavevmode
\epsfbox[200 450 400 720]{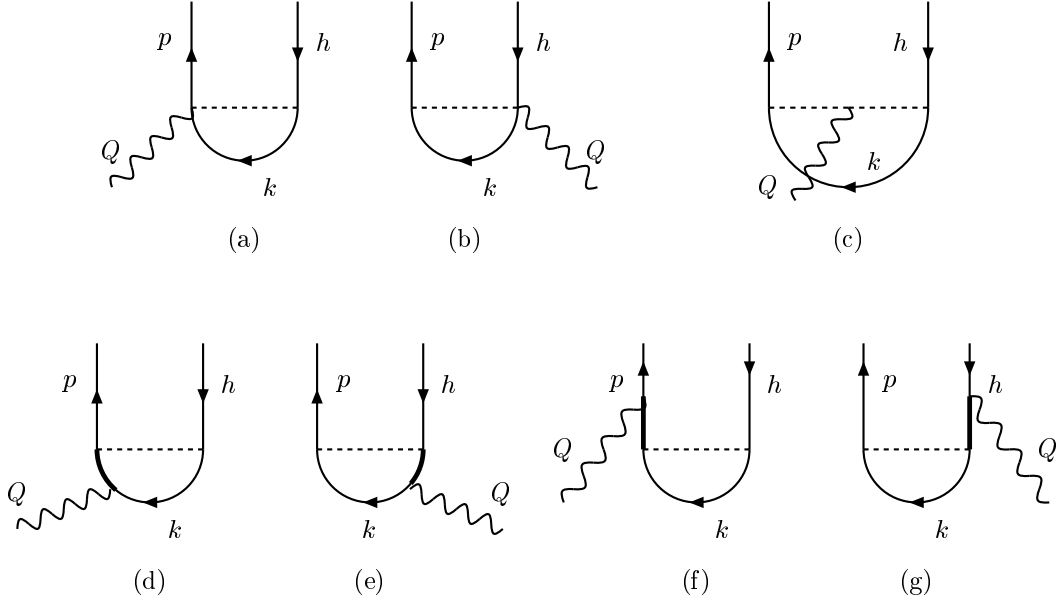}
\end{center}
\caption{Many-body Feynman diagrams contributing to the one-body
current with one pion-exchange. The thick line in the correlation
diagrams (d)--(g)
corresponds to a fully relativistic Dirac propagator for the
nucleon.  Diagrams (d)-(e) represent the vertex current, while
diagrams (f) and (g) represent the self-energy current of the
hole and of the particle, respectively.
}
\end{figure}

In what follows we restrict our attention to the correlation current
given by diagrams (d)--(g) of Fig.~2. 
Following the arguments presented
in~\cite{Ama01}, we split the total correlation current into a {\em
vertex} correlation current ($v.c.$, also referred to as ``exchange''
in~\cite{Alb90}, diagrams (d) and (e)) and a {\em self-energy}
correlation current ($s.e.$, diagrams (f) and (g)), according to
\begin{equation}
j_{cor}^{\mu}= j_{s.e.}^{\mu}+j_{v.c.}^{\mu}\ .
 \end{equation}
The vertex and self-energy currents then read~\cite{Ama01}
\begin{eqnarray}
j_{v.c.}^{\mu}(\np,\nh)
&=& 
-\frac{f^2}{Vm_\pi^2} 
\sum_{k<F}\frac{m}{E_\nk}
\overline{u}(\np)\tau_a\gamma_5(\Pbar-\Kbar)
\frac{u(\nk)\overline{u}(\nk)}{(P-K)^2-m_\pi^2}
\Gamma^{\mu}(Q)
S_F(K-Q)
\tau_a\gamma_5(\Pbar-\Kbar)u(\nh)
\nonumber\\
&-& 
\frac{f^2}{Vm_\pi^2} 
\sum_{k<F}\frac{m}{E_\nk}
\overline{u}(\np)
\tau_a\gamma_5(\Kbar-\Hslash)
S_F(K+Q)
\Gamma^{\mu}(Q)
\frac{u(\nk)\overline{u}(\nk)}{(K-H)^2-m_\pi^2}
\tau_a\gamma_5(\Kbar-\Hslash)u(\nh)
\nonumber\\
\label{vc}
\end{eqnarray}
and
\begin{eqnarray}
j_{s.e.}^{\mu}(\np,\nh)
&=& 
-\frac{f^2}{Vm_\pi^2}\sum_{k<F}\frac{m}{E_\nk}
\overline{u}(\np)
\tau_a\gamma_5(\Pbar-\Kbar)
\frac{u(\nk)\overline{u}(\nk)}{(P-K)^2-m_\pi^2}
\tau_a\gamma_5(\Pbar-\Kbar)
S_F(P)\Gamma^{\mu}(Q)u(\nh)
\nonumber\\
&-& 
\frac{f^2}{Vm_\pi^2}\sum_{k<F}\frac{m}{E_\nk}
\overline{u}(\nk)
\tau_a\gamma_5(\Kbar-\Hslash)
\frac{u(\nh)\overline{u}(\np)}{(K-H)^2-m_\pi^2}
\Gamma^{\mu}(Q)S_F(H)
\tau_a\gamma_5(\Kbar-\Hslash)u(\nk)\ , \nonumber \\
\label{self-energy-ph}
\end{eqnarray}
respectively
\footnote{Note that here, although the global factor $\displaystyle
\frac{m}{V\sqrt{E_pE_h}}$ has been extracted from the current, the
factor $\displaystyle \frac{m}{V E_k}$, associated with the internal
line, has to be retained inside the sum.}.

It is important to point out that in eq.~(\ref{self-energy-ph}) the
particle ($p$) and hole ($h$) are described in the Fermi gas by
unperturbed plane waves, {\it i.e.,} they are on-shell, and hence the
propagators $S_F(P)$ and $S_F(H)$ diverge.  On the one hand, this problem cannot
be cured at the level of the current matrix elements. On the other, 
we know that the full set of Feynman
diagrams (Fig.~2) is needed in order to get a consistent,
gauge-invariant, one-pion-exchange current and hence their presence
is required. Indeed, the
s.e. diagrams yield a far from negligible contribution to the nuclear
response functions as shown in~\cite{Ama01}, where they are computed
through the imaginary part of the first-order polarization
propagator. Hence the necessity of having to renormalize the
expression in eq.~(\ref{self-energy-ph}) arises.

This divergence of the diagrams
(f)-(g) is reminiscent of the well-known infinity occurring in standard
perturbative quantum field theory, when
self-energy insertions in the external legs are included in Feynman diagrams.
As is well-known, there one should renormalize the
theory by dressing the external legs, propagators and vertices. In the
nuclear matter case we assume that the particle-physics
effects are already accounted for by the
physical masses and electromagnetic form factors. However, an additional
nuclear physics renormalization that arises from the interaction of a nucleon
with the nuclear medium should be included at the one-pion-exchange 
level to account for the self-energy diagram.

The self-energy current in eq.~(\ref{self-energy-ph}) can be
written in the following form:
\begin{equation}\label{SEC}
j^{\mu}_{s.e.}(\np,\nh) =
\overline{u}(\np)\Sigma(P)S_F(P)\Gamma^{\mu}(Q)u(\nh)
+\overline{u}(\np)\Gamma^{\mu}(Q)S_F(H)\Sigma(H)u(\nh) \ ,
\end{equation}
where $\Sigma(P)$ is the nucleon self-energy matrix that in first order reads
\begin{equation}\label{SE1}
\Sigma(P) = 
-\frac{f^2}{Vm_\pi^2}\sum_k\frac{m}{E_\nk}
\tau_a\gamma_5(\Pbar-\Kbar)
\frac{u(\nk)\overline{u}(\nk)}{(P-K)^2-m_\pi^2}
\tau_a\gamma_5(\Pbar-\Kbar)\ .
\end{equation}
This is diagrammatically displayed in Fig.~3. The s.e.
matrix, shown in Fig.~3(c), corresponds
to the Fock term of the mean-field potential (the Hartree or direct term is
zero for pion exchange).

\begin{figure}[tb]
\begin{center}
\leavevmode
\epsfbox[200 450 400 720]{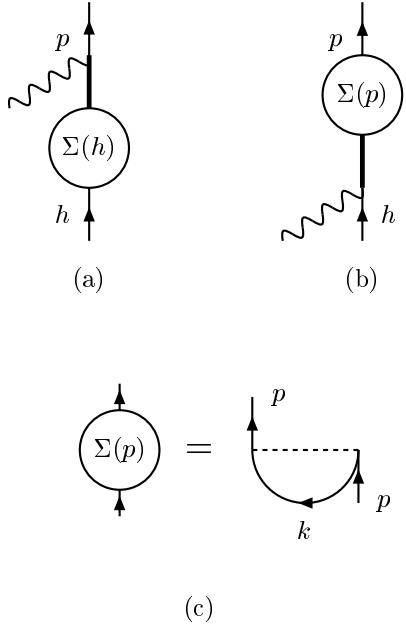}
\end{center}
\caption{ Diagrammatical representation of the self-energy current for
a hole (a) and a particle (b). The self-energy is defined to first
order as the Fock insertion shown in (c) with one pion-exchange.  }
\end{figure}

Performing the sum over the spin ($s$) and isospin ($t$) indices,
implicitly included in eq.~(\ref{SE1}), and using the commutation
properties of the gamma matrices to eliminate $\gamma_5$, the
self-energy can be cast in the form
\begin{equation}\label{SE2}
\Sigma(P) = -3\frac{f^2}{m_\pi^2}
\int\frac{d^3 k}{(2\pi)^3}\theta(k_F-k)
\frac{1}{2E_\nk}
\frac{(\Pbar-\Kbar)(\Kbar-m)(\Pbar-\Kbar)}{(P-K)^2-m_\pi^2}\ ,
\end{equation}
where the sum over $\nk$ has been converted into an integral.  Note
that the self-energies $\Sigma(P)$ and $\Sigma(H)$ appearing in
eq.~(\ref{SEC}) are evaluated for free particles and holes, {\it
i.e.}, for $P^{\mu}$ and $H^{\mu}$ on-shell. Hence the self-energy
contributions are divergent, since so are the free propagators $S(P)$
and $S(H)$ in eq.~(\ref{SEC}). Therefore they should not be computed
using eq.~(\ref{SEC}), but rather one should first renormalize the
wave function and the propagator of the particles in the medium. This
is achieved through the summation of the full series of diagrams with
repeated self-energy insertions displayed in Fig.~3.

Now the energy of a particle in nuclear matter is modified by the
interaction with the medium and, as well, through its energy-momentum
relation. Thus the associated momentum is no longer on-shell and
therefore in the next section we shall evaluate the self-energy for
{\em off-shell} particles. In the first iteration, corresponding to
one pion-exchange, the particle $P^{\mu}$ is off-shell, but the
intermediate interacting hole $K^{\mu}$ is not
\footnote{Note that in deriving eq.~(\ref{SE2}) we have assumed free
spinors $u(\nk)$; hence eq.~(\ref{SE2}) is only valid for $K^{\mu}$ 
on-shell. The off-shell case requires one to redefine the spinors
$u(\nk)$ according to an interacting Dirac equation, as is shown
later.}. In this case, with the help of Dirac spinology, one writes
\begin{equation}
(\Pbar-\Kbar)(\Kbar-m)(\Pbar-\Kbar)
= 
2(P\cdot K-m^2)(\Pbar+m) 
-(P^2-m^2)(\Kbar+m)\ ,
\end{equation}
which allows one to recast the self-energy in eq.~(\ref{SE2}) for the
off-shell momentum $P$ in the form
\begin{equation}\label{SE3}
\Sigma(P) = -3\frac{f^2}{m_\pi^2}
\int\frac{d^3 k}{(2\pi)^3}\theta(k_F-k)
\frac{1}{2E_\nk}
\frac{2(P\cdot K-m^2)(\Pbar+m)-(P^2-m^2)(\Kbar+m)}{(P-K)^2-m_\pi^2} \ .
\end{equation}
Note that the second term inside the integral vanishes for $P$ on-shell.

In general the self-energy of a nucleon in nuclear matter can be written in
the form~\cite{Cel86}:
\begin{equation}\label{spin}
\Sigma(P) = mA(P)+B(P)\gamma_0 p^0 -C(P)\ngamma\cdot\np \ .
\end{equation}
In contrast to the quantum field-theory decomposition $\Sigma(P)=
mA+B\Pbar$, owing to the non-invariance under a boost of the step
function $\theta(k_F-k)$ appearing in the self-energy, in nuclear
matter $B(P)\ne C(P)$. This in turn reflects the existence of a
privileged system, namely the lab system where the Fermi gas has ${\bf
p}_{FG}=0$, this being the natural one in which to compute the self-energy.
Under a boost, the Fermi gas ground state is no longer characterized
by $k<k_F$ and also the self-energy takes a different form.

In the case of the Fock self-energy in eq.~(\ref{SE3}) 
the functions $A,B,C$ can be expressed in terms of the integrals
\begin{eqnarray}
I(P) &\equiv&
\int\frac{d^3 k}{(2\pi)^3}\theta(k_F-k)
\frac{1}{2E_\nk}
\frac{1}{(P-K)^2-m_\pi^2}
\label{I}
\\
L^{\mu}(P) &\equiv&
\int\frac{d^3 k}{(2\pi)^3}\theta(k_F-k)
\frac{1}{2E_\nk}
\frac{K^{\mu}}{(P-K)^2-m_\pi^2}
\label{Lmu}
\end{eqnarray}
and in terms of the functions $K_0(P)$ and $K_3(P)$, defined 
as
\footnote{
${\bf L}$ is parallel to ${\bf p}$ since, choosing ${\bf p}$ along the $z$
axis, the azimuthal integration in eq.~(\ref{Lmu}) yields $L_x=L_y=0$.}
\begin{eqnarray}
L_0(P) &=& K_0(P)p_0 \\
{\bf L}(p) &=& K_3(P)\np\ .
\end{eqnarray}
Indeed one gets
\begin{eqnarray}
A(P) &=&
-3\frac{f^2}{m_\pi^2}
\left[ 
   2(P_\mu L^\mu(P) -m^2 I(P))- (P^2-m^2)I(P)
\right]
\label{A-definition}\\ 
B(P) &=&
-3\frac{f^2}{m_\pi^2}
\left[ 
   2(P_\mu L^\mu(P) -m^2 I(P))- (P^2-m^2)K_0(P)
\right]
\label{B-definition}\\ 
C(P) &=&
-3\frac{f^2}{m_\pi^2}
\left[ 
   2(P_\mu L^\mu(P) -m^2 I(P))- (P^2-m^2)K_3(P)
\right] \ .
\label{C-definition}
\end{eqnarray}
Note that $A=B=C$ for $P$ on-shell. In this case one has
\begin{equation}
\Sigma(P)_{on-shell}= A(P)(m+\Pbar)\ .
\end{equation}

\section{Hartree-Fock renormalization in nuclear matter}
\subsection{Nucleon Propagator}

The nucleon propagator in the nuclear medium results from
summing up a series with any number of self-energy insertions (see Fig.~4),
namely
\begin{figure}[tb]
\begin{center}
\leavevmode
\epsfbox[170 600 470 730]{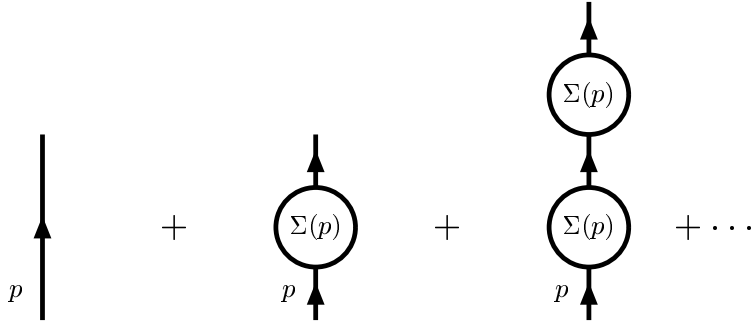}
\end{center}
\caption{
Diagrammatical series for the nucleon propagator in the medium.
}
\end{figure}
\begin{equation}
\frac{1}{\Pbar-m}
+
\frac{1}{\Pbar-m}
\Sigma(P)
\frac{1}{\Pbar-m}
+
\frac{1}{\Pbar-m}
\Sigma(P)
\frac{1}{\Pbar-m}
\Sigma(P)
\frac{1}{\Pbar-m}
+
\cdots
=
\frac{1}{\Pbar-m-\Sigma(P)}.
\end{equation}
Using the spin decomposition of the self-energy in eq.~(\ref{spin}), we
can write
\begin{equation}
\Pbar-m-\Sigma(P) 
= \left[1-B(P)\right]\gamma_0 p_0-
\left[1-C(P)\right]\ngamma\cdot\np -
\left[1+A(P)\right] m\ .
\end{equation}
Now the new four-momentum
$f^{\mu}=f^{\mu}(P)$, which is related to $P^{\mu}$ as follows
\begin{eqnarray}
f_0(P) &=& \frac{1-B(P)}{1-C(P)}\ p_0 \label{f_0}\\
{\bf f}(P) &=& \np \ ,
\end{eqnarray}
and the functions
\begin{eqnarray}
\widetilde{m}(P) &=& \frac{1+A(P)}{1-C(P)}\ m  \label{m-tilde}\\
z(P) &=& \frac{1}{1-C(P)}
\label{zp}
\end{eqnarray}
allow one to recast the nucleon propagator in the form 
\begin{equation} \label{propagator}
\frac{1}{\Pbar-m-\Sigma(P)}
= \frac{z(P)}{\gamma_0 f_0(P)-\ngamma\cdot\np -\widetilde{m}(P)}
= \frac{z(P)}{\fbar(P)-\widetilde{m}(P)}\ .
\end{equation}
For a nucleon with a fixed three-momentum $\np$, the pole
of the propagator in eq.~(\ref{propagator}) in the variable $p_0$ 
defines the new energy
of the nucleon in the medium. To find the latter
we introduce
\begin{equation}\label{Etilde}
\widetilde{E}(P)\equiv 
E(\np,\widetilde{m}(P))=\sqrt{\np^2+\widetilde{m}(P)^2}\ .
\end{equation}
Then the propagator reads
\begin{equation}
\frac{1}{\Pbar-m-\Sigma(P)}
= \frac{z(P)}{f_0(P)-\widetilde{E}(P)}
  \frac{\fbar(P)+\widetilde{m}(P)}{f_0(P)+\widetilde{E}(P)}
\label{propse}
\end{equation}
and its pole $p_0$ is found by solving the implicit equation 
\begin{equation}
f_0(P)=\widetilde{E}(P)\ ,
\end{equation}
which, exploiting eq.~(\ref{f_0}), can be recast as follows
\begin{equation}\label{dispersion}
p_0= \frac{1-C(P)}{1-B(P)}\sqrt{\np^2+\widetilde{m}(P)^2}
\equiv \frac{1-C(p_0,\np)}{1-B(p_0,\np)}
\sqrt{\np^2+\left[\widetilde{m}(p_0,\np)\right]^2}\ .
\label{p0eq}
\end{equation}
The solution of eq.~(\ref{p0eq}) for fixed $\np$ 
defines the new dispersion relation $p_0=\epsilon(\np)$
for interacting nuclear matter. 
Once the above equation has been solved,
the field strength renormalization constant
\begin{equation}
Z_2(\np)= {\rm Res}
\left. \frac{z(P)}{f_0(P)-\widetilde{E}(P)}\right|_{p_0=\epsilon(\np)}\ ,
\end{equation}
defined as the residue of the first factor on the right-hand side of
eq.~(\ref{propse}) at $p_0=\epsilon(\np)$, can be computed.
Indeed using eq.~(\ref{zp}), $Z_2(\np)$ is obtained 
by expanding the denominator around the pole $\epsilon(\np)$,
{\it i.e.,}
\begin{equation}
\left[1-C(P)\right]\left[f_0(P)-\widetilde{E}(P)\right]
= Z_2(\np)^{-1}\left[p_0-\epsilon(\np)\right]+ \cdots\ ;
\end{equation}
hence
\begin{eqnarray}
Z_2(\np)^{-1}
&=&\left.\frac{\partial}{\partial p_0}\right|_{p_0=\epsilon(\np)}
\left\{\left[1-C(P)\right]\left[f_0(P)-\widetilde{E}(P)\right]\right\}
\nonumber\\
&=& \left.\frac{\partial}{\partial p_0}\right|_{p_0=\epsilon(\np)}
\left\{\left[1-B(P)\right]p_0-\left[1-C(P)\right]\widetilde{E}(P)\right\}\ .
\end{eqnarray}
With the help of eq.~(\ref{Etilde}) 
the derivative can be easily evaluated, the result being
\begin{equation} \label{Z_2}
Z_2(\np)^{-1} = 
\left[ 1-B
-\frac{\partial B}{\partial p_0}p_0 
-m\frac{\widetilde{m}}{\widetilde{E}}\frac{\partial A}{\partial p_0} 
+\frac{\np^2}{\widetilde{E}}\frac{\partial C}{\partial p_0} 
\right]_{p_0=\epsilon(\np)}\ .
\end{equation}

\subsection{Nucleon Spinors}

The self-energy modifies not only the propagator and the
energy-momentum relation of a nucleon,
but, as well, the free Dirac spinors. In fact
the spinors are now solutions of the Dirac equation in
the nuclear medium~\cite{Cel86}, {\it i.e.},
\begin{equation}\label{new_spinors}
[\Pbar-m-\Sigma(P)]\phi(\np)=0\ ,
\end{equation}
which, again using the decomposition in eq.~(\ref{spin}),
can be recast as follows
\begin{equation}\label{Dirac}
\left[ \gamma_0 f_0(P)
       -\ngamma\cdot\np
       -\widetilde{m}(P)
\right]\phi(\np)=0\ ,
\end{equation}
the functions $f_0(P)$ and $\widetilde{m}(P)$ being
defined in eqs.~(\ref{f_0}) and (\ref{m-tilde}), respectively.
Equation~(\ref{Dirac}) has the same structure as the free Dirac equation;
hence for the positive-energy eigenvalue one has
\begin{equation}
f^2_0(P)= \np^2+\widetilde{m}^2(P)\ ,
\end{equation}
which implicitly yields the energy $p_0=\epsilon(\np)$ of the nucleon in
the nuclear medium. This result was already obtained as the
pole of the nucleon propagator. Then the corresponding 
positive-energy spinors ($s$ being the spin index) read
\begin{equation}
\phi_s(\np) = 
\sqrt{Z_2(\np)}
\left(
\frac{\widetilde{E}(\np)+\widetilde{m}(\np)}{2\widetilde{m}(\np)}
\right)^{1/2}
\columnmatrix{\chi_s}{
\frac{\nsigma\cdot\np}{\widetilde{E}(\np)+\widetilde{m}(\np)}\chi_s}
=
\sqrt{Z_2(\np)}u_s(\np,\widetilde{m}(\np))
\label{phispin}
\end{equation}
and the functions 
{\em of the three-momentum $\np$}
\be
\widetilde{m}(\np)
\equiv
\widetilde{m}(\epsilon(\np),\np)
\label{Dirac-mass}
\ee
\be
\widetilde{E}(\np) 
\equiv
\widetilde{E}(\epsilon(\np),\np) 
= \sqrt{\np^2+\widetilde{m}(\np)^2}
\label{Dirac-energy}
\ee
represent the nucleon's effective mass and effective
energy corresponding to $p_0=\epsilon(\np)$. The field strength
renormalization constant, $\sqrt{Z_2(\np)}$, of the new spinors, defined 
in eq.~(\ref{Z_2}), is required by renormalization theory, since
the propagator in eq.~(\ref{propagator}) for $p_0$ close to the
pole $\epsilon(\np)$ reads
\begin{equation}
\frac{1}{\Pbar-m-\Sigma(P)} 
\sim 
\frac{1}{p_0-\epsilon(\np)}
\frac{\widetilde{m}(\np)}{\widetilde{E}(\np)}
\sum_s \phi_s(\np)\overline{\phi}_s(\np)
=
\frac{Z_2(\np)}{p_0-\epsilon(\np)}
\frac{\fbar(\np)+\widetilde{m}(\np)}{2\widetilde{E}(\np)}\ .
\end{equation}
Once the new spinors have been computed, the self-energy can be
evaluated by inserting the latter into eq.~(\ref{SE1}). Then the Dirac
equation should be solved again with the new self-energy and so on.
This self-consistent procedure leads to the relativistic Hartree-Fock 
model which has to be dealt with numerically.

In this paper we do not attempt to solve the HF equations, since we
are interested only in the first-order correction to the
single-nucleon current. Although the latter cannot be derived by
directly applying the Feynman rules, it can still be identified with
the self-energy diagrams of Fig.~2 (f)--(g).  Thus in the next section
we shall compute the renormalized one-body current using the new
spinors and energy-momentum relation and then expand it in powers of
the square of the pion-coupling constant $f^2/m_\pi^2$.  As we shall
see, the unperturbed one-body current is thus recovered as the
leading-order term and the first-order one coincides with the
self-energy contribution.

\subsection{Non-relativistic self-energy}

Before performing the expansion of the renormalized
current we briefly examine the non-relativistic
limit of the self-energy diagrams in order to bring to light some
differences that exist with respect to the fully relativistic case. 

The non-relativistic leading order of the self-energy 
current in eq.~(\ref{self-energy-ph}) is obtained by using 
the following prescriptions, which are valid in the static limit,
\begin{eqnarray}
E_\nk &\simeq& m 
\\
\gamma_5\Kbar &\simeq& \nsigma\cdot\nk 
\\
\frac{1}{K^2-m_\pi^2} &\simeq& -\frac{1}{\nk^2+m_\pi^2}
\\
S_F(P) &\simeq& S_{nr}(P) = \frac{1}{p_0-\frac{\np^2}{2m}} \ .
\label{nr-propagator}
\end{eqnarray}
The electromagnetic form factor $\Gamma^\mu(Q)$ is also replaced by
$\Gamma^{\mu}_{nr}(q)$, representing
the usual non-relativistic one-body (OB)
current acting over bi-spinors~\cite{Ama98,Alb90}. 
Using the above relations and 
performing the sums over spin and isospin indices, the s.e. current matrix
element results
\begin{equation}
j^{\mu}_{se}(\np,\nh)
\simeq 
\chi_p^{\dagger}
\left[
       \Sigma_{nr}(\np)S_{nr}(P)\Gamma^{\mu}_{nr}(Q)
      +\Gamma^{\mu}_{nr}(Q)S_{nr}(H)\Sigma_{nr}(\nh)
\right]
\chi_h \ ,
\end{equation}
where $\chi_p$ and $\chi_h$ are two-components spinors.
The non-relativistic self-energy function is given by
\begin{equation} 
\label{nr-self-energy}
\Sigma_{nr}(\np)=3\frac{f^2}{Vm_\pi^2}\sum_{\nk}
\frac{(\np-\nk)^2}{(\np-\nk)^2+m_\pi^2}
=\Sigma_{nr}(|\np|) \ .
\end{equation}

The non-relativistic nucleon propagator in eq.~(\ref{nr-propagator}) 
in the medium is then 
\begin{equation}
S_{nr}(p_0,\np)=
 \frac{1}{p_0-\frac{\np^2}{2m}}
+\frac{1}{p_0-\frac{\np^2}{2m}}
 \Sigma_{nr}(\np)
 \frac{1}{p_0-\frac{\np^2}{2m}}
+\cdots = 
\frac{1}{p_0-\frac{\np^2}{2m}-\Sigma_{nr}(\np)}\ .
\label{nr-full-propagator}
\end{equation}
As is well-known, this is a meromorphic function whose simple pole 
again defines the new energy of the nucleon in the medium, namely
\begin{equation}\label{nr-energy}
\epsilon_{nr}(\np)=\frac{\np^2}{2m}+\Sigma_{nr}(\np)\ ,
\end{equation}
since $\Sigma_{nr}(\np)$ is a function only of $\np$.

Since the non-relativistic self-energy function in
eq.~(\ref{nr-self-energy}) does not depend on spin, the nucleon wave
functions are not modified in the medium. In fact the corresponding
Schr\"odinger equation in momentum space, including the self-energy,
is simply given by
\begin{equation}
\left[\frac{\np^2}{2m}+\Sigma_{nr}(\np)
\right] \phi_{nr}(\np) = p_0\phi_{nr}(\np)\ ,
\end{equation}
with the bi-spinor $\phi_{nr}(\np)$ corresponding to the eigenvalue 
$p_0=\epsilon_{nr}(\np)$. 

The non-relativistic analysis of the nucleon self-energy
current~\cite{FW} is much simpler than its relativistic
counterpart. Indeed, in the former the self-consistency is immediately
achieved because the nucleon wave functions are not modified by the
self-energy interaction and thus the first iteration of the
``Hartree-Fock'' equations already provides the exact energy.
Instead, in the relativistic framework the spin dependence of the
self-energy modifies the Dirac-spinors, inducing an enhancement of the
lower components. Moreover, the field-strength renormalization
constant, namely the residue of the nucleon propagator in
eq.~(\ref{nr-full-propagator}) at the pole, in the non-relativistic
case is just one.  Hence the enhancement of the lower components and
the spinors' field strength renormalization are genuine relativistic
effects that are absent in a non-relativistic analysis where only the
energy-momentum relation in the medium is altered by the self-energy
diagrams.  We shall show in next section that the two above-mentioned
relativistic signatures can be recast as new pieces in the
electromagnetic current acting over free spinors.

\section{Self-energy current to first order}

The particle-hole current matrix element in the HF approximation reads
\begin{equation}\label{HF-current}
j_{HF}^{\mu}(\np,\nh)= \overline{\phi}(\np)\Gamma^{\mu}(Q)\phi(\nh) \ ,
\end{equation}
where $\phi(\np)$ are the new renormalized HF spinors introduced in
eq.~(\ref{new_spinors}). Hence eq.~(\ref{HF-current}) represents the
electromagnetic excitation of the p-h pair with dressed external lines
corresponding to the sum of the diagrams shown in Fig.~5.
\begin{figure}[tb]
\begin{center}
\leavevmode
\epsfbox[160 580 570 750]{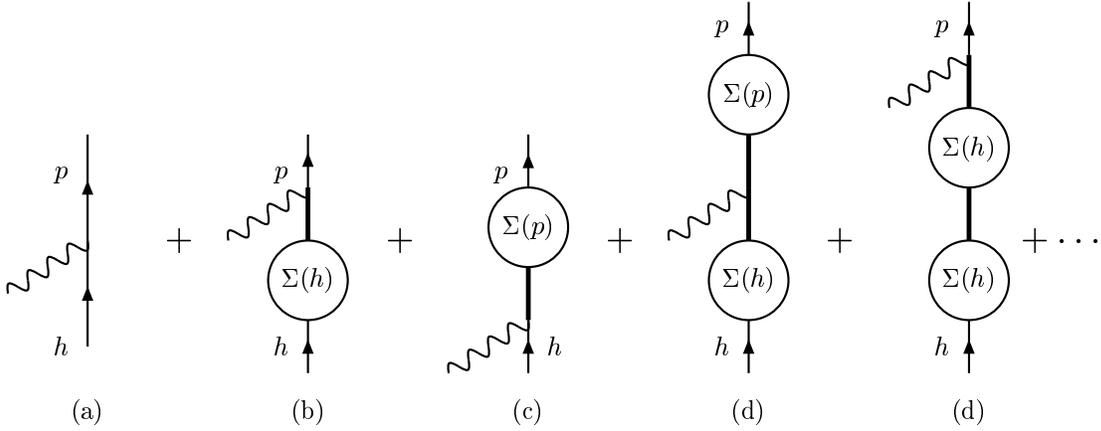}
\end{center}
\caption{
Diagrammatical series for the one-body electromagnetic current with 
dressed external lines.
}
\end{figure}
In order to obtain a one-pion-exchange expression we expand
eq.~(\ref{HF-current}) in powers of the square of the pion coupling
constant $f^2/m_\pi^2$ and single out the first-order term, that is,
the one that is linear in $f^2/m_\pi^2$. We shall still refer to the
current thus obtained that represents the OPE contribution as the
``self-energy'' current and, importantly, we shall show that it yields
a finite contribution, free from the divergence problem of the current
in eq.~(\ref{SEC}).

To proceed we start by deriving the HF energy $\epsilon(\np)$ to first
order in $f^2/m_\pi^2$.  For this purpose we note that the functions
$A(P)$, $B(P)$ and $C(P)$ defined in
eqs.~(\ref{A-definition}--\ref{C-definition}) are of order
$O(f^2/m_\pi^2)$.  Hence the following expansion of the Dirac mass in
eq.~(\ref{m-tilde}) holds:
\begin{equation}\label{m-expansion-1}
\widetilde{m}(P)=m\frac{1+A(P)}{1-C(P)}
= m\left[1+A(P)+C(P)\right]+O(f^4/m_\pi^4)\ .
\end{equation}
Inserting this into eq.~(\ref{dispersion}) for the energy
and expanding again to first order in $f^2/m_\pi^2$, we get
\begin{eqnarray}
p_0=\epsilon(\np)
&\simeq& 
[1-C(P)+B(P)]\sqrt{\np^2+m^2+2m^2\left[A(P)+C(P)\right]}
\nonumber\\
&=& 
E_\np+\Delta E(p_0,\np)\ ,
\end{eqnarray}
where $E_\np= \sqrt{\np^2+m^2}$ is the unperturbed free energy and
\begin{eqnarray}
\Delta E(p_0,\np)\equiv 
\frac{1}{E_\np}(m^2 A+E_\np^2 B-\np^2 C) +O(f^4/m_\pi^4)
\label{energycorrection}
\end{eqnarray}
is the first-order correction to the energy.
Next we can insert the above value of $p_0$ 
{\em inside} the argument of the functions $A$, $B$, $C$.
Expanding the latter around the on-shell value $p_0=E_\np$ we get
\begin{equation}
A(P)=A(p_0,\np)=A(E_\np+\Delta E,\np)= 
A(E_\np,\np)+O(f^4/m_\pi^4)
\simeq A_0(\np)\ ,
\end{equation}
where $A_0(\np) \equiv A(E_p,\np)$.
Likewise to first order we obtain 
\begin{eqnarray}
B(P) &\simeq & B(E_\np,\np)\equiv B_0(\np) \\
C(P) &\simeq & C(E_\np,\np)\equiv C_0(\np) \ .
\end{eqnarray}
Recalling that for $P$ on-shell
the functions $A$, $B$, $C$ coincide, {\it i.e.,}
$A_0(\np)= B_0(\np)= C_0(\np)$,
we can insert these on-shell values 
into eq.~(\ref{energycorrection}) and, neglecting terms of second order,
{\it i.e.,} $O(f^4/m_\pi^4)$, we finally arrive at the result
\begin{equation}
p_0=\epsilon(\np)= E_\np+\frac{1}{E_\np}B_0(\np)(m^2+E_\np^2-\np^2)
+O(f^4/m_\pi^4)=
E_\np+\frac{1}{E_\np}2m^2B_0(\np)
+O(f^4/m_\pi^4)\ .
\label{HFE}
\end{equation}
The above expression can be recast in terms of the
on-shell value of the self-energy 
\begin{equation}\label{Sigma_0}
\Sigma_0(\np)\equiv 2mB_0(\np)\ ,
\end{equation}
which satisfies the relation
\begin{equation}
\Sigma(E_\np,\np)u(\np) = \Sigma_0(\np)u(\np)\ ,
\end{equation}
thus showing that the free spinors are eigenvectors of the self-energy matrix 
$\Sigma(E_\np,\np)$ corresponding to the  eigenvalue $\Sigma_0(\np)$.
Hence to first order in $f^2/m_\pi^2$, the HF energy in eq.~(\ref{HFE})
is found to read
\begin{equation}\label{energy}
\epsilon(\np)
\simeq 
E_\np+ \frac{m}{E_\np}\Sigma_0(\np)
\end{equation}
in terms of the on-shell self-energy eigenvalue $\Sigma_0(\np)$. 
When compared with the
non-relativistic HF energy in eq.~(\ref{nr-energy}) it appears that, beyond
the different expressions of the self-energy functions that hold in
the relativistic and non-relativistic frameworks, an extra 
multiplicative factor $m/E_\np$ occurs in the relativistic case.

Once the HF energy $\epsilon(\np)$ is known to first order
in $f^2/m_\pi^2$, we expand as well the renormalized
spinors, namely
\begin{equation}\label{ren-spinor}
\sqrt{\frac{\widetilde{m}(\np)}{\widetilde{E}(\np)}}
u(\np,\widetilde{m}(\np))
= 
\sqrt{\frac{\widetilde{E}(\np)+\widetilde{m}(\np)}{2\widetilde{E}(\np)}}
\columnmatrix{\chi}%
{\frac{\nsigma\cdot\np}{\widetilde{E}(\np)+\widetilde{m}(\np)}\chi}\ .
\end{equation}
Actually, for later use in the calculation of the hadronic tensor, it is
convenient to expand the spinor multiplied by the factor
$\sqrt{\frac{\widetilde{m}(\np)}{\widetilde{E}(\np)}}$. 

Thus we start by expanding the
Dirac mass in eq.~(\ref{m-expansion-1}) around the on-shell energy,
obtaining
\begin{equation}
\widetilde{m}(\np)
=
m\left[1+A_0(\np)+C_0(\np)\right]+O(f^2/m_\pi^2)
\simeq  m+\Sigma_0(\np)\ ,
\label{m-expansion}
\end{equation}
where use has been made of the on-shell self-energy in eq.~(\ref{Sigma_0}).
Likewise, using the HF equation (eq.~(\ref{dispersion})),
the Dirac energy $\widetilde{E}(\np)$ defined in eq.~(\ref{Dirac-energy})
is given by
\begin{eqnarray}
\widetilde{E}(\np)
&=& \frac{1-B}{1-C}\epsilon(\np)
\simeq
\left[1-B_0(\np)+C_0(\np)\right]\left[E_\np+\frac{m}{E_\np}\Sigma_0(\np)
\right]
\nonumber\\
&\simeq& E_\np+ \frac{m}{E_\np}\Sigma_0(\np)
\simeq \epsilon(\np)\ .
\label{E-expansion}
\end{eqnarray}
After some algebra the following first-order expressions are obtained
\begin{eqnarray}
\sqrt{\frac{\widetilde{E}+\widetilde{m}}{2\widetilde{E}}}
&\simeq&
\sqrt{\frac{m+E_\np}{2E_\np}}
\left( 1+\frac{E_\np-m}{2E_\np}\frac{\Sigma_0}{E_\np}\right)
\label{norm}\\
\frac{1}{\widetilde{E}+\widetilde{m}}
&\simeq&
\frac{1}{m+E_\np}\left(1-\frac{\Sigma_0}{E_\np}\right)\ .
\label{1Em}
\end{eqnarray}
Inserting eqs.~(\ref{norm}) and~(\ref{1Em}) into the renormalized spinor
in eq.~(\ref{ren-spinor}) we get
\begin{eqnarray}
\sqrt{\frac{\widetilde{m}(\np)}{\widetilde{E}(\np)}}
u(\np,\widetilde{m}(\np))
&\simeq&
\sqrt{\frac{m+E_\np}{2E_\np}}
\left[ 1+\frac{E_\np-m}{2E_\np}\frac{\Sigma_0}{E_\np}\right]
\columnmatrix{\chi}%
{\frac{\nsigma\cdot\np}{m+E_\np}\left(1-\frac{\Sigma_0}{E_\np}\right)\chi}
\nonumber\\
&\simeq&
\sqrt{\frac{m}{E_\np}}u(\np)
+
\frac{\Sigma_0}{E_\np}
\sqrt{\frac{m}{E_\np}}
\sqrt{\frac{m+E_\np}{2m}}
\columnmatrix{\frac{E_\np-m}{2E_\np}\chi}%
{-\frac{E_\np+m}{2E_\np}\frac{\nsigma\cdot\np}{m+E_\np}\chi}\ .
\end{eqnarray}
Since
\begin{equation}
\columnmatrix{(E_\np-m)\chi}%
{-(E_\np+m)\frac{\nsigma\cdot\np}{m+E_\np}\chi}
=
(E_\np\gamma_0-m)
\columnmatrix{\chi}{\frac{\nsigma\cdot\np}{m+E_\np}\chi}\ .
\end{equation}
the first-order (in $f^2/m_\pi^2$) renormalized spinor 
can be cast in the form
\begin{equation}
\sqrt{\frac{\widetilde{m}(\np)}{\widetilde{E}(\np)}}
u(\np,\widetilde{m}(\np))
\simeq
\sqrt{\frac{m}{E_\np}}
\left[ 
      u(\np)
     +\frac{\Sigma_0(\np)}{E_\np}\frac{E_\np\gamma_0-m}{2E_\np}u(\np)
\right]\ .
\end{equation}

We now also expand the field-strength renormalization 
function defined in eq.~(\ref{Z_2}). For this purpose we use
eqs.~(\ref{m-expansion},\ref{E-expansion}), 
obtaining
\begin{equation}
Z_2(\np) \simeq 
\left[ 1+B_0(\np)
+\frac{m^2}{E_\np}\frac{\partial A}{\partial p_0} 
+E_\np\frac{\partial B}{\partial p_0}
-\frac{\np^2}{E_\np}\frac{\partial C}{\partial p_0} 
\right]_{p_0=E_\np}\ ,
\end{equation}
which implies that
\begin{equation}
\sqrt{Z_2(\np)} \simeq 1+\frac12\alpha(\np)
\end{equation}
with
\begin{equation} \label{alpha}
\alpha(\np) \equiv
 B_0(\np)
+\left[\frac{m^2}{E_\np}\frac{\partial A}{\partial p_0} 
+E_\np\frac{\partial B}{\partial p_0}
-\frac{\np^2}{E_\np}\frac{\partial C}{\partial p_0} 
\right|_{p_0=E_\np}\ .
\end{equation}

Finally, collecting the above results and inserting them into 
eq.~(\ref{phispin}), we get to first order 
\begin{equation}\label{expanded-spinor}
\sqrt{\frac{\widetilde{m}(\np)}{\widetilde{p_0}(\np)}}
\phi(\np)
\simeq 
\sqrt{\frac{m}{E_\np}}
\left[ 
      u(\np)
     +\frac{\Sigma_0}{E_\np}\frac{E_\np\gamma_0-m}{2E_\np}u(\np)
     +\frac12\alpha(\np) u(\np)
\right]\ .
\end{equation}
Thus, within the OPE approach the renormalized HF spinors in the
nuclear medium are characterized by two new elements with respect to
the bare $u(\np)$. In appendix A we study in detail the second term on
the right-hand side of eq.~(\ref{expanded-spinor}). We show that
$(E_p\gamma_0-m)u(\np)$ is directly connected with the negative-energy
components in the wave function. Thus the second term in
eq.~(\ref{expanded-spinor}) involves
$\frac{\Sigma_0}{E_p}\frac{E_p\gamma_0-m}{2E_p}u(\np)$, which can be
written as a difference between two divergent terms, one being
$S_F(\np)\Sigma(\np)u(\np)$ and the other a kind of
``renormalization'' counterterm of the original self-energy current.
The final result turns out, however, to be a {\em finite} quantity.

Using the above expressions for the renormalized spinors, 
we now expand the renormalized one-body current matrix element
to first order in $f^2/m_\pi^2$, getting
\begin{eqnarray}
  \sqrt{\frac{\widetilde{m}(\np)}{\widetilde{E}(\np)}}
  \sqrt{\frac{\widetilde{m}(\nh)}{\widetilde{E}(\nh)}}
  j^{\mu}_{HF}(\np,\nh)
&\simeq&
\sqrt{\frac{m}{E_\np}}
\sqrt{\frac{m}{E_\nh}}
\overline{u}(\np)
\left[ 
      \Gamma^{\mu}
     +\Gamma^{\mu}
       \frac{\Sigma_0(\nh)}{E_\nh}\frac{E_\nh\gamma_0-m}{2E_\nh}
     +\frac{\alpha(\nh)}{2}
      \Gamma^{\mu}
\right.
\nonumber\\
&&
\kern 6em
\left.\mbox{}
     +\frac{\Sigma_0(\np)}{E_\np}\frac{E_\np\gamma_0-m}{2E_\np}
      \Gamma^{\mu}
     +\frac{\alpha(\np)}{2}
      \Gamma^{\mu}
\right]
u(\nh)
\nonumber\\
&\equiv&
\frac{m}{\sqrt{E_\np E_\nh}}
\left[ j^{\mu}_{OB}(\np,\nh)+j^{\mu}_{RSE}(\np,\nh) \right]\ .
\label{HF-current-expansion}
\end{eqnarray}
In eq.~(\ref{HF-current-expansion}) the term $j^{\mu}_{OB}$
represents the usual one-body current matrix element evaluated 
with free spinors, {\it i.e.,}
\begin{equation}
j^{\mu}_{OB}(\np,\nh)= \overline{u}(\np)\Gamma^{\mu}(Q)u(\nh) \ ,
\end{equation}
whereas $j^{\mu}_{RSE}$ is a new renormalized self-energy current matrix 
element
that includes the effects of the renormalization of the spinors. 
It can be decomposed according to
\begin{equation}\label{RSE}
j^{\mu}_{RSE}(\np,\nh)
= j^{\mu}_{RSE1}(\np,\nh)+j^{\mu}_{RSE2}(\np,\nh)\ ,
\end{equation}
where  $j^{\mu}_{RSE1}$ embodies the correction arising from
the new spinor solution of the Dirac equation in the medium
and $j^{\mu}_{RSE2}$ the one stemming from the field-strength 
renormalization function $\sqrt{Z_2}$ in the medium. 
Their expressions are the following:
\begin{eqnarray} 
j^{\mu}_{RSE1}(\np,\nh)
&=&
\overline{u}(\np)
\left[ 
     \Gamma^{\mu}
       \frac{\Sigma_0(\nh)}{E_\nh}\frac{E_\nh\gamma_0-m}{2E_\nh}
     +\frac{\Sigma_0(\np)}{E_\np}\frac{E_\np\gamma_0-m}{2E_\np}
      \Gamma^{\mu}
\right]
u(\nh)
\label{RSE1}
\\
j^{\mu}_{RSE2}(\np,\nh)
&=&
\left[\frac{\alpha(\nh)+\alpha(\np)}{2}\right]j^{\mu}_{OB}(\np,\nh)\ .
\label{RSE2}
\end{eqnarray}

\subsection{Gauge Invariance}

As shown above, the renormalized HF current matrix element, expanded
to first order in $f^2/m_\pi^2$, has been split into the usual
one-body current and into a new renormalized self-energy current.  In
order to be consistent with the one-pion-exchange model, we should add
the contribution of the seagull, pion-in-flight and vertex correlation
currents corresponding to the
diagrams shown in Fig.~2(a--e).  We point out once more that the
self-energy diagrams (f) and (g), corresponding to insertions in
external legs, should be disregarded in computing amplitudes (or
currents) in perturbation theory. Instead, their contributions should
be taken into account via renormalized energies and spinors as
solutions of the relativistic HF equations. We have expressed the
latter, to first order in $f^2/m_\pi^2$, in the form of a new current
operator (denoted as RSE current).

Then the total current in our model reads
\begin{equation}\label{jfull1}
j^{\mu}(\np,\nh)= j^{\mu}_{OB}(\np,\nh)+ j^{\mu}_{OPE}(\np,\nh)\ ,
\label{jtotal}
\end{equation}
where $j^{\mu}_{OPE}$ embodies the seagull, pion-in-flight, vertex
correlation and renormalized self-energy currents, namely
\begin{equation}
j^{\mu}_{OPE}= j^{\mu}_{s}+ j^{\mu}_{p}+j^{\mu}_{v.c.}+j^{\mu}_{RSE}\ .
\label{jfull}
\end{equation}
In what follows we shall prove the gauge invariance of this current to
first order in $f^2/m_\pi^2$. In so-doing it is crucial to take into
account not only the full current in eqs.~(\ref{jfull1},\ref{jfull}),
but also the first-order correction to the energy of the particles and
holes due to the self-energy interaction in eq.~(\ref{energy}).  In
other words, for a given momentum transfer $\nq=\np-\nh$, the energy
transfer should be computed as the difference between the particle and
hole HF energies and not using the free values $E_\np$ and $E_\nh$.
Hence the energy transfer is
\begin{equation}
\omega_{HF} = E_\np-E_\nh +
\frac{m}{E_\np}\Sigma_0(\np)-\frac{m}{E_\nh}\Sigma_0(\nh)
\end{equation}
and, in conformity, the associated four-momentum transfer is
$Q_{HF}^{\mu}=(\omega_{HF},\nq)$. To make the following discussion clearer 
we denote with
$Q_{HF}$ the HF four-momentum and with  $\omega_{HF}$ the HF  energy 
transfer, to 
distinguish them from the on-shell values $Q$ and $\omega$.

\paragraph{\bf \underline{Divergence of the one-body current}\\}

The divergence of the zeroth-order one-body current computed using the
HF four-momentum transfer $Q_{HF}$ is given by
\begin{equation}
Q_{HF,\mu}j^{\mu}_{OB}(\np,\nh)
= \overline{u}(\np)Q_{HF,\mu}
\Gamma^{\mu}\left( Q_{HF} \right) u(\nh)
= \overline{u}(\np)F_1\left(Q_{HF}\right)\Qbar_{HF}u(\nh)\ ,
\end{equation}
where the nucleon vertex $\Gamma^{\mu}(Q_{HF})$ is evaluated at the 
momentum transfer $Q_{HF}$. 
Since $\overline{u}(\np)\!\!\!\Qbar u(\nh)=0$, only
the first-order contribution arising from the self-energy
correction survives, namely
\begin{equation}  \label{OB-div}
Q_{HF,\mu}j^{\mu}_{OB}(\np,\nh)
= \overline{u}(\np)F_1(Q)
\left[\frac{m}{E_\np}\Sigma_0(\np)-\frac{m}{E_\nh}\Sigma_0(\nh)\right]
\gamma_0 u(\nh)\ .
\end{equation}
In the above the Dirac form factor $F_1$ is computed at
the unperturbed value $Q^{\mu}$, since we disregard second-order
contributions.
Note that the one-body current itself is not gauge
invariant --- its divergence yields a first-order term which 
turns out to be essential for the gauge invariance of the full current,
as we shall see below.

\paragraph{\bf \underline{Divergence of the MEC}\\}

As shown in~\cite{Ama01}, after
summing over the intermediate spin and isospin indices,
the following seagull and pion-in-flight 1p-1h matrix elements
\begin{eqnarray}
j_{s}^{\mu}(\np,\nh)
&=& 
-\frac{f^2}{Vm_\pi^2} 
F_1^V i \epsilon_{3ab}
\overline{u}(\np)\tau_a\tau_b
\sum_{\nk}\frac{m}{E_\nk}
\left\{
\frac{(\Kbar-m)\gamma^{\mu}}{(P-K)^2-m_\pi^2}
+\frac{\gamma^{\mu}(\Kbar-m)}{(K-H)^2-m_\pi^2}
\right\}
u(\nh)
\nonumber\\
\label{seagull}\\
j_{p}^{\mu}(\np,\nh)
&=& 
\frac{f^2}{Vm_\pi^2} 
F_1^V i \epsilon_{3ab}
\overline{u}(\np)\tau_a\tau_b
\sum_{\nk}\frac{m}{E_\nk}
\frac{2m(Q+2H-2K)^{\mu}}{[(P-K)^2-m_\pi^2] [(K-H)^2-m_\pi^2]}
u(\nh)
\nonumber\\
\label{pion-in-flight}
\end{eqnarray}
are obtained.
These currents are already of first order in $f^2/m_\pi^2$; thus in
computing their divergence one 
must use the unperturbed value of the energy
transfer~\cite{Ama01}, otherwise a second-order term arises.
Then using the free Dirac equation
and exploiting the kinematics we obtain (see~\cite{Ama01} for 
details)
\begin{eqnarray}
Q_{\mu}j_{s}^{\mu}(\np,\nh)
&=& 
-\frac{f^2}{Vm_\pi^2} 
F_1^V i \epsilon_{3ab}
\overline{u}(\np)\tau_a\tau_b
\sum_{\nk}\frac{m}{E_\nk}
\left\{
\frac{2(K\cdot P-m\Kbar)}{(P-K)^2-m_\pi^2}
-\frac{2(K\cdot H-m\Kbar)}{(K-H)^2-m_\pi^2}
\right\}
u(\nh)
\nonumber\\
\label{seagull-div}\\
Q_{\mu}j_{p}^{\mu}(\np,\nh)
&=& 
-\frac{f^2}{Vm_\pi^2} 
F_1^V i \epsilon_{3ab}
\overline{u}(\np)\tau_a\tau_b
\sum_{\nk}\frac{m}{E_\nk}
\left\{
\frac{2m(\Kbar-m)}{(P-K)^2-m_\pi^2}
-\frac{2m(\Kbar-m)}{(K-H)^2-m_\pi^2}
\right\}
u(\nh)\ .
\nonumber\\
\label{pion-in-flight-div}
\end{eqnarray}
Upon addition of these two equations the terms containing $\Kbar$ cancel,
leaving for the total divergence of the seagull and pion-in-flight
the expression
\begin{equation}
Q_{\mu}(j_{s}^{\mu}+j_p^{\mu})= 
-\frac{f^2}{Vm_\pi^2} 
F_1^V i \epsilon_{3ab}
\overline{u}(\np)\tau_a\tau_b
\sum_{\nk}\frac{2m}{2E_\nk}
\left\{
\frac{2(K\cdot H-m^2)}{(K-H)^2-m_\pi^2}
-\frac{2(K\cdot P-m^2)}{(P-K)^2-m_\pi^2}
\right\}
u(\nh)\ ,
\end{equation}
which can be further simplified by introducing
the self-energy in eq.~(\ref{SE3}) for on-shell momenta, yielding finally
\begin{equation}
Q_{\mu}(j_{s}^{\mu}+j_p^{\mu})= 
\frac{i}{3} F_1^V  \epsilon_{3ab}
\overline{u}(\np)\tau_a\tau_b
[\Sigma(\np)-\Sigma(\nh)]u(\nh)\ .
\label{MEC-div}
\end{equation}

\paragraph{\bf \underline{Divergence of the vertex correlation current}\\}

Starting from the 1p-1h matrix element of the v.c. current in eq.~(\ref{vc})
and applying the Dirac equation, we get (see~\cite{Ama01} for details)
\begin{eqnarray}
Q_{\mu}j_{v.c.}^{\mu}(\np,\nh)
&=& 
\frac{f^2}{Vm_\pi^2} 
\overline{u}(\np)\tau_a F_1\tau_a 
\sum_k\frac{1}{2E_\nk}
\gamma_5(\Pbar-\Kbar)
\frac{\Kbar+m}{(P-K)^2-m_\pi^2}
\gamma_5(\Pbar-\Kbar)
u(\nh)
\nonumber\\
&-& 
\frac{f^2}{Vm_\pi^2} 
\overline{u}(\np)
\tau_a F_1 \tau_a
\sum_k\frac{1}{2E_\nk}
\gamma_5(\Kbar-\Hslash)
\frac{\Kbar+m}{(K-H)^2-m_\pi^2}
\gamma_5(\Kbar-\Hslash)u(\nh)\ ,
\nonumber\\
\end{eqnarray}
where we recognize again the expression of the self-energy 
matrix in eq.~(\ref{SE2}). 
Since the Dirac form factor can be split into an isoscalar and an 
isovector component according to
\begin{equation}
F_1 = \frac12(F_1^S+F_1^V\tau_3)\ ,
\end{equation}
which yields
\begin{equation}
\tau_a F_1 \tau_a = 3 F_1 +iF_1^V\epsilon_{3ab}\tau_a\tau_b\ ,
\end{equation}
the divergence of the v.c. current written
in terms of the self-energy function reads
\begin{equation}
Q_{\mu}j_{v.c.}^{\mu}(\np,\nh) = 
\overline{u}(\np)
\left(
       F_1 + \frac{i}{3}F_1^V\epsilon_{3ab}\tau_a\tau_b
\right)
\left[\Sigma(\nh)-\Sigma(\np)\right] u(\nh)\ .
\end{equation}
Comparing this result with eq.~(\ref{MEC-div}) we note that
the $\epsilon$ term above cancels with the MEC 
contribution. Hence 
\begin{equation}\label{Ward}
Q_{\mu}\left[j_{MEC}^{\mu}(\np,\nh)+j_{v.c.}^{\mu}(\np,\nh)\right]
= \overline{u}(\np)F_1\left[\Sigma(\nh)-\Sigma(\np)\right]u(\nh)\ .
\end{equation}
The above relation just expresses the Ward-Takahashi identity relating the
full vertex correction, namely MEC plus v.c.
(diagrams 2 (a)--(e)), to the self-energy matrix element.  

\paragraph{\bf \underline{Divergence of the RSE current}\\}

Here we obtain the divergence of the renormalized self-energy (RSE)
current defined in eqs.~(\ref{RSE},\ref{RSE1},\ref{RSE2}). 
For this purpose we first note that the divergence 
of $j^\mu_{RSE2}$ vanishes to first order because it is
proportional to the OB current. Hence we write
\begin{equation} 
Q_{\mu}j^{\mu}_{RSE}(\np,\nh)
=
\overline{u}(\np)
\left[ 
     F_1\Qbar
       \frac{\Sigma_0(\nh)}{E_\nh}\frac{E_\nh\gamma_0-m}{2E_\nh}
     +\frac{\Sigma_0(\np)}{E_\np}\frac{E_\np\gamma_0-m}{2E_\np}
     F_1\Qbar
\right]
u(\nh).
\end{equation}
Using the relation 
$\overline{u}(\np)\Qbar u(\nh)=0$ and  
\begin{eqnarray}
\overline{u}(\np)\Qbar\gamma_0 u(\nh)
&=&
\overline{u}(\np)2(m\gamma_0-E_\nh) u(\nh)
\\
\overline{u}(\np)\gamma_0\Qbar u(\nh)
&=&
\overline{u}(\np)2(E_\np-m\gamma_0) u(\nh)
\end{eqnarray}
it is straightforward to obtain
\begin{equation}
Q_{\mu}j^{\mu}_{RSE}(\np,\nh) =
\overline{u}(\np)F_1\left[\Sigma(\np)-\Sigma(\nh)\right] u(\nh)
+\overline{u}(\np)F_1
\left[\frac{m}{E_\nh}\Sigma_0(\nh)-\frac{m}{E_\np}\Sigma_0(\np)\right] u(\nh)\ .
\end{equation}

Remarkably the first term of this equation cancels with the divergence
of the MEC plus the v.c. current, given by the Ward-Takahashi identity
in eq.~(\ref{Ward}), whereas the second term cancels with the
divergence of the OB current in eq.~(\ref{OB-div}). We have thus
proven that, within the present model up to first order in
$f^2/m_\pi^2$, the total current in eq.~(\ref{jtotal}) satisfies the
continuity equation, namely
\begin{equation}
Q_{HF,\mu}(j_{OB}^{\mu}+j_{MEC}^{\mu}+j_{v.c.}^{\mu}+j_{RSE}^{\mu})=0\ .
\end{equation}

\section{Relativistic self-energy responses}

In this section we derive the self-energy contribution to the nuclear
response functions to first order in $f^2/m_\pi^2$.  We can thus
compare with the results obtained in~\cite{Ama01} using the
polarization propagator formalism. Our goal is to show that the
results obtained in the two formalisms coincide, although they stem
from different approaches, as we have previously emphasized.

The one-body hadronic tensor in HF approximation reads
\begin{equation}
W^{\mu\nu}_{HF}(\omega,\nq)
= \sum_{s_ps_h}\sum_{t_pt_h}
\int\frac{d^3 h}{(2\pi)^3}
\frac{\widetilde{m}(\np)\widetilde{m}(\nh)}%
{\widetilde{E}(\np)\widetilde{E}(\nh)}
 j_{HF}^{\mu}(\np,\nh)^* j_{HF}^{\nu}(\np,\nh)
\delta(\omega+\epsilon(\nh)-\epsilon(\np))\ ,
\end{equation}
where $\np=\nh+\nq$ and $j_{HF}^{\mu}(\np,\nh)$ is the one-body HF 
current in eq.~(\ref{HF-current})
computed using the renormalized HF spinors and HF
energies of the particle and the hole.

Next we use the expansions in eqs.~(\ref{HF-current-expansion}) 
for the current $j_{HF}$ and (\ref{energy}) for the HF energies.
In addition we expand the energy delta function to first order  
in $f^2/m_\pi^2$ according to
\begin{equation}
\delta(\omega+\epsilon(\nh)-\epsilon(\np))
\simeq 
\delta(\omega+E_\nh-E_\np)
+\frac{d \delta(\omega+E_\nh-E_\np)}{d\omega}
\left[
       \frac{m}{E_\nh}\Sigma_0(\nh)
      -\frac{m}{E_\np}\Sigma_0(\np)
\right]\ .
\end{equation}
Inserting all of these relations into the hadronic tensor and
neglecting terms of second order we get for the diagonal elements of
the hadronic tensor
\footnote{We only work out the diagonal elements of the hadronic tensor,
since these are the ones that contribute to the unpolarized inclusive 
longitudinal and transverse response functions.}
\begin{equation}
W^{\mu\mu}_{HF}(\omega,\nq)
\simeq 
W^{\mu\mu}_{OB}(\omega,\nq)
+\Delta W^{\mu\mu}_{RSE}(\omega,\nq)
\label{Wmumu}
\end{equation}
(the summation convention is not in force in eq.~(\ref{Wmumu})),
where $W^{\mu\mu}_{OB}(\omega,\nq)$ is the usual OB hadronic tensor of
a RFG, {\it i.e.},
\begin{equation}
W^{\mu\mu}_{OB}=\sum_{s_ps_h}\sum_{t_pt_h}
\int\frac{d^3h}{(2\pi)^3}
\frac{m^2}{E_\np E_\nh}
       |j^{\mu}_{OB}(\np,\nh)|^2
\delta(\omega+E_\nh-E_\np)\ ,
\end{equation}
and $\Delta W^{\mu\mu}_{RSE}(\omega,\nq)$ is the first-order
self-energy correction 
\begin{eqnarray}
\Delta W^{\mu\mu}_{RSE}
&=&
\sum_{s_ps_h}\sum_{t_pt_h}
\int\frac{d^3h}{(2\pi)^3}
\frac{m^2}{E_\np E_\nh}
\left\{ \phantom{\frac12}
        2{\Re}\; j^{\mu}_{OB}(\np,\nh)^*j^{\mu}_{RSE}(\np,\nh)
        \delta(\omega+E_\nh-E_\np)
\right.
\nonumber\\
&& \mbox{}+\left.
       |j^{\mu}_{OB}(\np,\nh)|^2
\left[
       \frac{m}{E_\nh}\Sigma_0(\nh)
      -\frac{m}{E_\np}\Sigma_0(\np)
\right]
       \frac{d}{d\omega} \delta(\omega+E_\nh-E_\np)
\right\}\ .
\label{RSE-response}
\end{eqnarray}
In eq.~(\ref{RSE-response}) the first term corresponds to the interference 
between the OB and the RSE
currents, while the second one, which shifts the
allowed kinematical region because of the derivative 
of the energy delta function, is due to the modification of the
nucleon energies in the medium.

Carrying out the spin traces for the single-nucleon current 
\begin{equation}
\sum_{s_ps_h}
|j^{\mu}_{OB}(\np,\nh)|^2
=
\frac{1}{4m^2}
{\rm Tr}
\left\{
     \Gamma^{\mu}(Q)
(\Hslash+m)\Gamma^{\mu}(-Q)(\Pbar+m)
\right\} \ ,
\end{equation}
we get for the renormalized self-energy response function 
\begin{eqnarray}
\Delta W^{\mu\mu}_{RSE} 
&=&
\int\frac{d^3h}{(2\pi)^3}
\frac{1}{4E_\np E_\nh}
{\rm Tr}
\left\{
     \Gamma^{\mu}(Q)
\left[ 
       \frac{\Sigma_0(\nh)}{E_\nh}\frac{E_\nh\gamma_0-m}{2E_\nh}
      +\frac{\alpha(\nh)}{2}
\right]
(\Hslash+m)\Gamma^{\mu}(-Q)(\Pbar+m)
\right.
\nonumber\\
&&
\kern 3cm
+
\Gamma^{\mu}(Q)
(\Hslash+m)\Gamma^{\mu}(-Q)(\Pbar+m)
\left[ 
     \frac{\Sigma_0(\np)}{E_\np}\frac{E_\np\gamma_0-m}{2E_\np}
+
\frac{\alpha(\np)}{2}
\right]
\nonumber\\
&&
\kern 3cm
+
  \Gamma^{\mu}(Q)
  (\Hslash+m)
  \left[ 
        \frac{\Sigma_0(\nh)}{E_\nh}\frac{E_\nh\gamma_0-m}{2E_\nh}
       +\frac{\alpha(\nh)}{2}
  \right]
  \Gamma^{\mu}(-Q)
  (\Pbar+m)
\nonumber\\
&&
\kern 3cm
+
\left.
 \Gamma^{\mu}(Q)
 (\Hslash+m)
 \Gamma^{\mu}(-Q)
 \left[ 
     \frac{\Sigma_0(\np)}{E_\np}\frac{E_\np\gamma_0-m}{2E_\np}
     +\frac{\alpha(\np)}{2}
 \right]
 (\Pbar+m)
\right\}
\nonumber\\
&&        
\kern 3cm
\times
\delta(\omega+E_\nh-E_\np)
\nonumber\\
&& \mbox{}+
\int\frac{d^3h}{(2\pi)^3}
\frac{1}{4E_\np E_\nh}
{\rm Tr}
\left\{
     \Gamma^{\mu}(Q)
(\Hslash+m)\Gamma^{\mu}(-Q)(\Pbar+m)
\right\}
\nonumber\\
&&
\kern 3cm
\left(
       \frac{m}{E_\nh}\Sigma_0(\nh)
      -\frac{m}{E_\np}\Sigma_0(\np)
\right)
       \frac{d}{d\omega} \delta(\omega+E_\nh-E_\np)\ .
\label{W-RSE}
\end{eqnarray}
More precisely, one should add two copies of eq.~(\ref{W-RSE}),
one with the form factors appropriate to the proton and one to the neutron.

In appendix B we show that this contribution to the response function
is identical to the one obtained in~\cite{Ama01} computing the
imaginary part of the polarization propagator corresponding to the two
diagrams of Fig.~6.
\begin{figure}[tb]
\begin{center}
\leavevmode
\epsfbox[160 540 470 700]{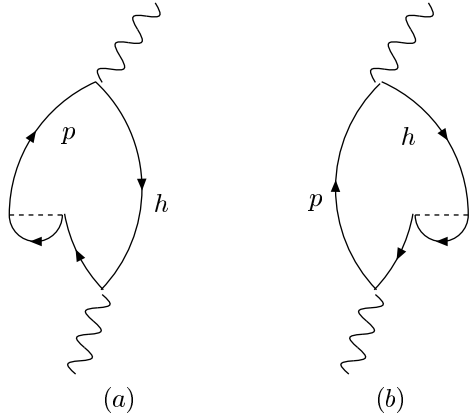}
\end{center}
\caption{First-order contributions to the 
polarization propagator with self-energy insertions in the
hole or particle lines.
}
\end{figure}
This identity is not trivial: indeed in the case of the polarization
propagator the response functions, with the Fock self-energy dressing
the particle and the hole lines, are computed by representing the
product of two nucleon propagators as the derivative of a single one
to deal with the presence of a double pole in the integrand.  In the
present paper the problem has been solved differently. First the
entire perturbative series with Fock self-energy insertions has been
summed up and then the result has been expanded to first order, thus
obtaining a finite first-order current operator. Because of the
equivalence of these two procedures (as proven in appendix B) we are
confident about the validity of the results we have obtained for the
self-energy contribution to the nuclear responses.

Finally we display the results obtained in our framework
for the nuclear response functions\footnote{For clarity the results
presented here have been calculated ignoring the hadronic
($\pi NN$) form factors.}. 
In particular, we explore the impact on the latter of
the new currents $j^\mu_{RSE1}$ and $j^\mu_{RSE2}$ that arise from
the enhancement of the lower components of the spinors and from the 
field strength renormalization $\sqrt{Z_2(\np)}$, respectively.

In Fig.~7 we show the on-shell self-energy (solid curve) and the
field strength renormalization function (dashed curve) given by 
eqs.~(\ref{Sigma_0}) and (\ref{alpha}), respectively. The explicit 
expressions for $\Sigma_0(p)$ and $\alpha(p)$ are derived in Appendix C.
The $\Sigma_0(p)$ obtained here is in good agreement with the results of
ref.~\cite{Ana81} and its effect on the single-particle energy in
eq.~(\ref{energy})
and on the effective mass in eq.~(\ref{m-expansion}) is very small (less than 
$\sim 3\%$). 
Note that $\alpha$, which is 
linked to the current $j^\mu_{RSE2}$  of eq.~(\ref{RSE2}),
is much smaller than $\Sigma_0(p)/E_\np$, which enters in $j^\mu_{RSE1}$
through eq.~(\ref{RSE1}). 
This implies that the effect of the enhancement of the
lower components of the spinors dominates over the field-strength 
renormalization. This is very clearly seen in Figs.~8 and 9, where the various
contributions to the longitudinal and transverse responses stemming from 
$j^\mu_{RSE1}$ and  $j^\mu_{RSE2}$ are displayed versus the transferred energy
$\omega$ for momentum transfer $q=$ 0.5, 1, 2 and 3 GeV/c.
It is evident that the effect of  $j^\mu_{RSE2}$ is negligible with respect to
that of  $j^\mu_{RSE1}$.
The separate contributions of the particle and hole self-energies are also 
shown:
as $q$ increases the contribution of the particle is suppressed, whereas the
one for the hole survives. 

In Fig.~10 we compare the contribution to the longitudinal response
of the renormalization of the wave 
functions (dashed) with that arising from the renormalization of the 
energies (solid). It clearly appears that 
the effect that is linked to modification of the energy due to the medium
is the dominant one, the other being very small, especially for large 
values of $q$.
Similar results are obtained for the transverse response.

\begin{figure}[tb]
\begin{center}
\leavevmode 
\input{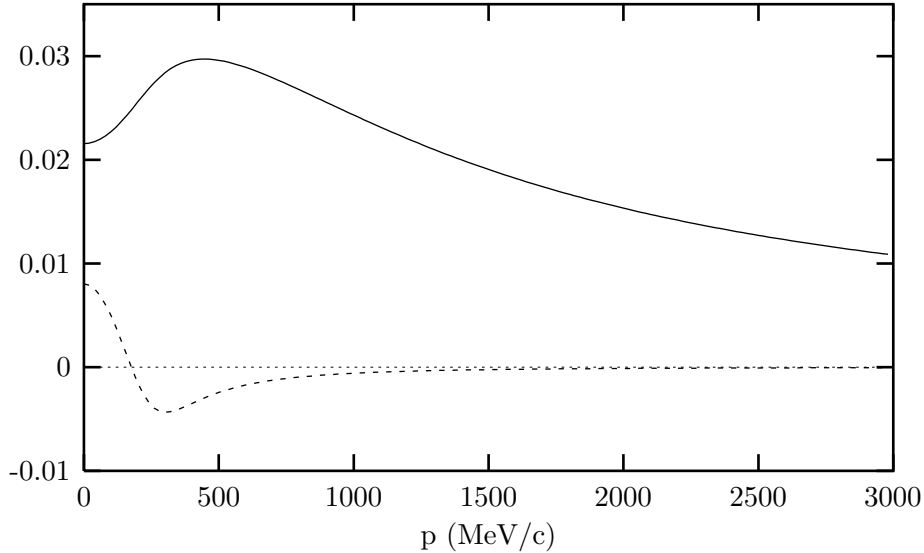}
\end{center} 
\caption{The on-shell self-energy $\Sigma_0(p)/E_\np$ defined
in eq.~(\protect\ref{Sigma_0}) (solid line) and the field-strength 
renormalization
function $\alpha(p)$ given in eq.~(\protect\ref{alpha}) 
(dashed line) plotted versus the
momentum $p$.}
\end{figure}

\begin{figure}[tb]
\begin{center}
\leavevmode
\input{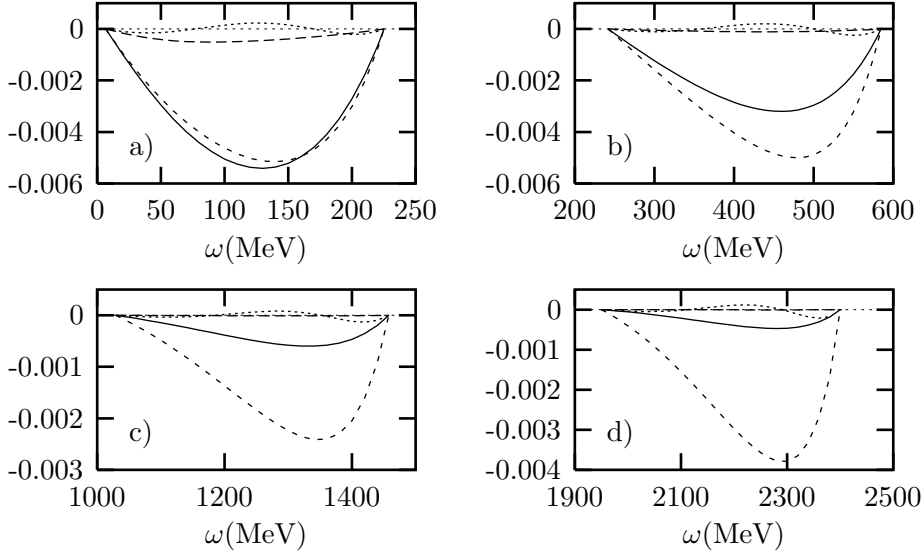}
\end{center}
\caption{The contribution of the renormalized self-energy current to the
longitudinal response plotted versus $\omega$. The nucleus is $^{40}$Ca with 
$k_F$=237 MeV/c and the units are 
$10^{-1}$ MeV$^{-1}$ at $q$=0.5 GeV/c (panel a),
$10^{-2}$ MeV$^{-1}$ at $q$=1 GeV/c (panel b),
$10^{-3}$ MeV$^{-1}$ at $q$=2 GeV/c (panel c),
$10^{-4}$ MeV$^{-1}$ at $q$=3 GeV/c (panel d).
The separate contributions of the current $j^\mu_{RSE1}$ for the particle
(solid) and hole (short-dashed) and of the current $j^\mu_{RSE2}$ for the 
particle (long-dashed) and hole (dotted) are displayed. 
}
\end{figure}

\begin{figure}[bt]
\begin{center}
\leavevmode
\input{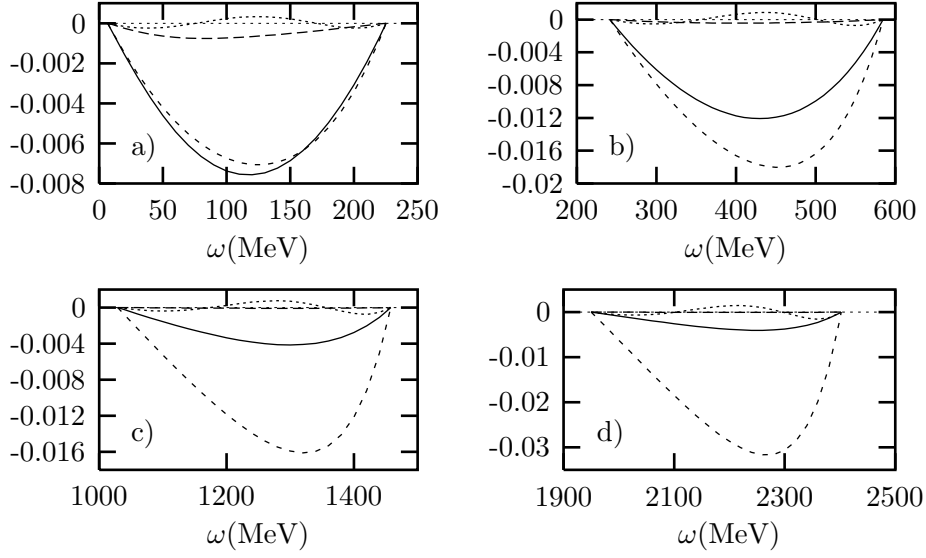}
\end{center}
\caption{Same as Fig.~8 for the transverse response.}
\end{figure}

\begin{figure}[tb]
\begin{center}
\leavevmode
\input{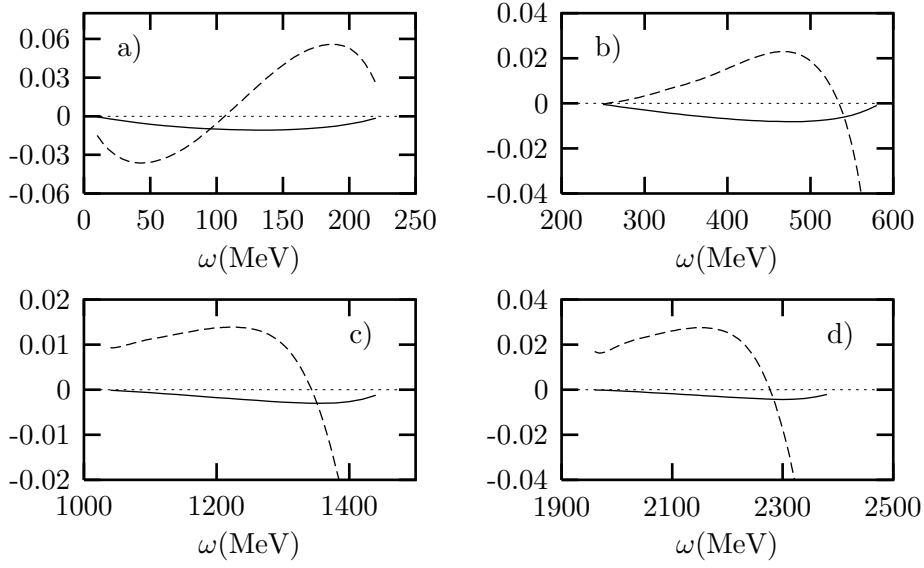}
\end{center}
\caption{The contributions of the first (solid) and second (dashed) term in 
eq.~(\protect\ref{RSE-response}) to the longitudinal response is plotted.
The units and labelling are as in Fig.~8.}
\end{figure}

\section{Conclusions}

As a more in-depth extension of the analysis carried out
in~\cite{Ama01}, in this paper we have again studied the set of 
one-pion-exchange operators that contribute to the electromagnetic
responses of nuclei in the 1p-1h channel. Our main concern in this
systematic approach has been to preserve fundamental symmetries such
as Lorentz covariance and gauge invariance.  As a zeroth-order
approximation we have used the RFG, this model being motivated by its
ability to handle issues of covariance and gauge invariance in a
reasonably transparent manner. Its limitations are not expected to be
too severe, since our entire focus is placed on the kinematic region
of the QEP~\cite{Ama94b}.  To go beyond the RFG implies a
consideration of interactions between the nucleons, and as a first
step, in this work we have limited the scope to one pion-exchange,
postponing for future study a more complete treatment including
heavier mesons.

In assessing the role of the pions in the electromagnetic nuclear
responses, the MEC are not the only contributions that arise in
first-order perturbation theory. In fact the pionic correlations are
intimately linked to MEC through the continuity equation and only when
the full set of Feynmam diagrams with one pion-exchange is considered
can one expect gauge invariance to be fulfilled.

Concerning such correlations, in this paper we have paid special
attention to the self-energy contribution. Indeed the iteration of the
self-energy diagrams generates a ``dressed'' propagator in the medium.
By the same token the self-energy generates ``dressed'' or
``renormalized'' wave functions in the medium, solutions of an
in-medium Dirac equation, where the self-energy plays the role of a
mean relativistic potential.  This equation also provides the
dispersion relation linking the energy and momentum of the nucleon in
the medium. Importantly, the new spinors should be multiplied by a
renormalization function $\sqrt{Z_2(\np)}$.

As the self-energy is generated by the interaction of a nucleon with
the other nucleons in the medium, the solutions of the new Dirac
equation should be used as input to re-compute the self-energy and so
on. The exact answer is obtained through a self-consistent procedure.
In this paper, however, we have just considered the first iteration:
we have thus computed the self-energy current confining ourselves to
first-order corrections to the energy and spinors -- or, equivalently,
to corrections linear in the self-energy -- which correspond to
diagrams with only one pionic line, in order to be consistent with MEC
and the vertex correlation currents.

Notably in the first-order expansion of the renormalized spinors two
new elements with respect to the non-relativistic approach emerge, one
arising from the negative-energy components in the wave function
produced by the interaction, and the second from the renormalization
function $\sqrt{Z_2}$. These two corrections can be expressed as a new
renormalized self-energy current, $j^{\mu}_{RSE}$, acting over free
spinors. Our results show that, for typical kinematics, the former constitutes
a correction to the total self-energy of roughly 10--20\%, whereas the
latter is negligible. Moreover, while at low momentum transfers both
particle and hole contributions in the former play a role, at high $q$
only the hole contribution remains.

\subsection*{Acknowledgements}
J.E.A. wants to thank J. Nieves for useful discussions. 
This work was partially supported by funds provided by DGICYT (Spain) 
under Contracts Nos. PB/98-1111, PB/98-0676 and PB/98-1367 and the Junta
de Andaluc\'{\i}a (Spain), by the Spanish-Italian Research Agreement
HI1998-0241, by the ``Bruno Rossi'' INFN-CTP Agreement, by the 
INFN-CICYT exchange and in part by the U.S. Department of Energy under
Cooperative Research Agreement No. DE-FC02-94ER40818.

\section*{Appendix A. Properties of the spinor $u(\np,\widetilde{m}(\np))$}

The expansion 
\begin{equation}
\sqrt{\frac{\widetilde{m}(\np)}{\widetilde{E}(\np)}}
u(\np,\widetilde{m}(\np))
\simeq
\sqrt{\frac{m}{E}}
\left[
      u(\np)
     +\frac{\Sigma_0}{E}\frac{E\gamma_0-m}{2E}
      u(\np)
\right]
\label{A1}
\end{equation}
transparently displays the effect of the self-energy on the free spinor
$u(\np)$.
Indeed the second term in the square brackets of eq.~(\ref{A1}) corresponds to a
negative-energy component with momentum $\np$. In fact, the Dirac
equation for a positive-energy spinor is given by
\begin{equation}
(\np\cdot\ngamma+m)u(\np)=E\gamma_0 u(\np),
\kern 1cm
\mbox{with $E>0$} \ .
\end{equation}
Now if we apply the operator $(\np\cdot\ngamma+m)$ to the spinor
$(E\gamma_0-m)u(\np)$ we obtain
\begin{eqnarray}
(\np\cdot\ngamma+m)(E\gamma_0-m)u(\np)
&=&
\np\cdot\ngamma(E\gamma_0-m)u(\np)
+m(E\gamma_0-m)u(\np)
\nonumber\\
&=&
(-E\gamma_0-m)\np\cdot\ngamma u(\np)
+m(E\gamma_0-m)u(\np)
\nonumber\\
&=&
(-E\gamma_0-m)(E\gamma_0-m) u(\np)
+m(E\gamma_0-m)u(\np)
\nonumber\\
&=&
(-E\gamma_0-m+m)(E\gamma_0-m) u(\np)
\nonumber\\
&=&
-E\gamma_0(E\gamma_0-m) u(\np).
\label{A129}
\end{eqnarray}
Hence $(E\gamma_0-m)u$ is an eigenvalue of the free Dirac Hamiltonian
with eigenvalue $-E$. Therefore the operator $E\gamma_0-m$ transforms a
positive-energy spinor $u(\np)$ into a negative-energy one.

Moreover, it is useful to write down the spinor correction in an alternative
form. Using the identity in eq.~(\ref{A129}) we can write
\begin{equation}
(\Pbar-m)(E\gamma_0-m)u(\np)=2E\gamma_0(E\gamma_0-m) u(\np) \ .
\end{equation}
Dividing by $2E(\Pbar-m)$ we then obtain
\begin{equation}
\frac{E\gamma_0-m}{2E}u(\np)
= \frac{1}{\Pbar-m}\gamma_0(E\gamma_0-m) u(\np) \ .
\end{equation}
Hence the second term in the square brackets of the right-hand 
side of eq.~(\ref{A1})
can be recast in the form
\begin{eqnarray}
\frac{\Sigma_0}{E}
\frac{E\gamma_0-m}{2E}
u(\np)
&=&
\frac{\Sigma_0}{E}\frac{1}{\Pbar-m}\gamma_0(E\gamma_0-m)u(\np)
=
\frac{1}{\Pbar-m}
\left(1-\frac{m}{E}\gamma_0\right)
\Sigma(\np)u(\np)
\nonumber\\
&=&
S_F(\np)\left(1-\frac{m}{E}\gamma_0\right)\Sigma(\np)u(\np) \ .
\label{A132}
\end{eqnarray}
The first term in eq.~(\ref{A132}), $S_F(\np)\Sigma(\np)u(\np)$, 
corresponds to the one that enters in the original (divergent) 
self-energy current for a nucleon on-shell (eq.~(\ref{SEC})). 
The subtracted term, with the factor $\frac{m}{E}\gamma_0$ inserted between 
the propagator
and self-energy, cancels the divergence and yields a finite result.
Thus it can be viewed
as a ``recipe'' to renormalize the self-energy current.

\section*{Appendix B. Renormalized self-energy  
response using the polarization propagator}

In~\cite{Ama01} we computed the first-order self-energy 
contribution to the polarization propagator corresponding 
to the two diagrams of Fig.~6.
The latter splits into the sum of the two terms, with 
Fock self-energy insertions in the hole and particle lines
respectively, and reads
\begin{eqnarray}
\Pi^{\mu\nu}
&=&
-i {\rm Tr}\int\frac{d^4h}{(2\pi)^4}
\left\{ \Gamma^{\mu}(Q)S(H)\Sigma(H)S(H)\Gamma^{\nu}(-Q)S(H+Q)
\right.
\nonumber\\
&&
\kern 2.5cm
\mbox{}\left.
       +\Gamma^{\mu}(Q)S(H)\Gamma^{\nu}(-Q)S(H+Q)\Sigma(H+Q) S(H+Q)
\right\} \ ,
\label{Pi}
\end{eqnarray}
where $S(K)$ is the nucleon propagator in the medium, namely
\begin{eqnarray}
S(K) 
&=& (\Kbar+m)
    \left[\frac{1}{K^2-m^2+i\epsilon}
         + 2\pi i \theta(k_F-k)
        \delta(K^2-m^2)\theta(k_0)
    \right] 
\nonumber\\
&=& (\Kbar+m)
      \left[ \frac{\theta(k-k_F)}{K^2-m^2+i\epsilon}
            +\frac{\theta(k_F-k)}{K^2-m^2-i\epsilon k_0}
      \right] \ .
\end{eqnarray}
As in \cite{Alb90} the double poles appearing in the integrand
of eq.~(\ref{Pi}) are treated by writing 
the product of the two propagators as a derivative. Thus for the hole
propagators one has
\begin{eqnarray}
S(H)\Sigma(H)S(H) 
&=&
(\Hslash+m)\Sigma(H)
(\Hslash+m)
      \left[ \frac{\theta(h-k_F)}{(H^2-m^2+i\epsilon)^2}
            +\frac{\theta(k_F-h)}{(H^2-m^2-i\epsilon h_0)^2}
      \right]
\nonumber\\
&&\kern -3cm 
=
(\Hslash+m)\Sigma(H)
(\Hslash+m)
\left.\frac{d}{d\alpha}\right|_{\alpha=0}
\left[
       \frac{1}{H^2-\alpha-m^2+i\epsilon}
      + 2\pi i
        \theta(k_F-h)
        \delta(H^2-\alpha-m^2)\theta(h_0) 
\right]
\nonumber\\
\end{eqnarray}
and likewise for the propagators of the particle, $S(P)$.
This technique, however, cannot be applied to the divergent 
self-energy current in eq.~(\ref{SEC}), and accordingly we had 
to invoke renormalization techniques.

Using the above prescription,
subtracting the contribution coming from the vacuum and taking the imaginary
part, the following expression was obtained for the self-energy
contribution to the hadronic tensor~\cite{Ama01}:
\begin{eqnarray}
W^{\mu\nu}
&=&
\left.\frac{d}{d\alpha}\right|_{\alpha=0}
\int\frac{d^3h}{(2\pi)^3}
\frac{I^{\mu\nu}_{10}(E'_\nh(\alpha),\nh;E_\np,\np;q)}
{4E'_\nh (\alpha)E_\np}
\delta(E'_\nh (\alpha)+q_0-E_\np)
        \theta(k_F-h)
        \theta(p-k_F)
\nonumber\\
&+&
\left.\frac{d}{d\beta}\right|_{\beta=0}
\int\frac{d^3h}{(2\pi)^3}
\frac{I^{\mu\nu}_{01}(E_\nh,\nh;E'_\np(\beta),\np;q)}
{4E_hE'_\np(\beta)}
\delta(E_\nh+q_0-E'_\np(\beta))
        \theta(k_F-h)
        \theta(p-k_F) \ ,
\nonumber\\
\label{W-se}
\end{eqnarray}
where $\np=\nh+\nq$, and modified energies for holes
and particles, namely
\begin{eqnarray}
E'_\nh(\alpha) &=& \sqrt{\nh^2+\alpha+m^2}
\\
E'_\np(\beta)  &=& \sqrt{\np^2+\beta+m^2} \ ,
\end{eqnarray}
have been introduced (in the above $\alpha$ and $\beta$ are real parameters).
Finally the functions $I^{\mu\nu}_{nm}$ 
are so defined 
\begin{eqnarray}
I^{\mu\nu}_{10}(H,P,Q)
&=&
{\rm Tr}
[\Gamma^{\mu}(Q)(\Hslash+m)\Sigma(H)(\Hslash+m)\Gamma^{\nu}(-Q)(\Pbar+m)]
\\
I^{\mu\nu}_{01}(H,P,Q)
&=&
{\rm Tr}
[\Gamma^{\mu}(Q)(\Hslash+m)\Gamma^{\nu}(-Q)(\Pbar+m)\Sigma(P)(\Pbar+m)] \ .
\label{I01}
\end{eqnarray}

In order to prove the equivalence between the responses computed using the
polarization propagator in eq.~(\ref{W-se}) and the result 
in eq.~(\ref{W-RSE}),
obtained using the renormalized current and energies,
we perform the derivative with respect to $\alpha$ and $\beta$. For a general
function $F(h_0)$ we have
\begin{equation}
\left.\frac{dF(E'_\nh(\alpha))}{d\alpha}\right|_{\alpha=0}= 
\frac{1}{2E_\nh}\left[\frac{dF(h_0)}{dh_0}\right]_{h_0=E_\nh} \ .
\end{equation}
Hence, interchanging the derivatives and the integral,
we get for the hadronic tensor the expression
\begin{eqnarray}
W^{\mu\nu}
&=&
\int\frac{d^3h}{(2\pi)^3}
\frac{1}{4E_\nh E_\np}
\frac{d}{d h_0}
\left[\frac{I^{\mu\nu}_{10}(h_0,\nh;E_\np,\np;q)}{2h_0}\right]_{h_0=E_\nh}
\delta(E_\nh+q_0-E_\np)
        \theta(k_F-h)
        \theta(p-k_F)
\nonumber\\
&+&
\int\frac{d^3h}{(2\pi)^3}
\frac{1}{4E_\nh E_\np}
I^{\mu\nu}_{10}(E_\nh,\nh;E_\np,\np;q)
\frac{1}{2E_\nh}
\frac{d}{d q_0}
\delta(E_\nh+q_0-E_\np)
        \theta(k_F-h)
        \theta(p-k_F)
\nonumber\\
&+&
\int\frac{d^3h}{(2\pi)^3}
\frac{1}{4E_\nh E_\np}
\frac{d}{d p_0}
\left[\frac{I^{\mu\nu}_{01}(E_\nh,\nh;p_0,\np;q)}{2p_0}\right]_{p_0=E_\np}
\delta(E_\nh+q_0-E_\np)
        \theta(k_F-h)
        \theta(p-k_F)
\nonumber\\
&-&
\int\frac{d^3h}{(2\pi)^3}
\frac{1}{4E_\nh E_\np}
I^{\mu\nu}_{01}(E_\nh,\nh;E_\np,\np;q)
\frac{1}{2E_\np}
\frac{d}{d q_0}
\delta(E_\nh+q_0-E_\np)
        \theta(k_F-h)
        \theta(p-k_F) \ . \nonumber \\
\end{eqnarray}
In differentiating the function $I^{\mu\nu}_{01}$ in eq.~(\ref{I01}), we
first consider the term:
\begin{eqnarray}
\lefteqn{\frac{d}{dp_0}
\left[\frac{1}{2p_0}(\Pbar+m)\Sigma(P)(\Pbar+m)\right]_{p_0=E_\np}}
\\
&=&
\frac{\Sigma_0(\np)}{2E_\np}
\left[
-\frac{2m}{E_\np}(\Pbar+m)
+\gamma_0(\Pbar+m)
+(\Pbar+m)\gamma_0
\right]_{p_0=E_\np}
\nonumber\\
&&
+
\left[
\frac{1}{2E_\nh}(\Hslash+m)
\frac{\partial\Sigma(H)}{\partial p_0}
(\Hslash+m)
\right]_{p_0=E_\np} \ ,
\label{derivative}
\end{eqnarray}
where use has been made of the results
\begin{eqnarray}
\Sigma(P)(\Pbar+m) &=&\Sigma_0(\np)(\Pbar+m)
\label{spm}
\\
(\Pbar+m)(\Pbar+m) &=&  2m(\Pbar+m) \ ,
\label{pm2}
\end{eqnarray}
which hold for $P^\mu$ on-shell and
where $\Sigma_0(\np)$ is the eigenvalue of the self-energy 
for on-shell spinors.

Next we should compute the derivative of the self-energy
$\Sigma(P)$. This function has the general structure given in eq.~(\ref{spin}),
and its derivative implies derivatives of the coefficients $A$, $B$,
and $C$, namely
\begin{equation}
\frac{\partial\Sigma(P)}{\partial p_0}
=
m\frac{\partial A(P)}{\partial p_0}
+\frac{\partial B(P)}{\partial p_0} \gamma_0 p_0 
-\frac{\partial C(P)}{\partial p_0} \ngamma\cdot\np
+B(P)\gamma_0 \ ,
\end{equation}
which must be evaluated for $P^{\mu}$ on-shell. 
Using again eq.~(\ref{pm2}) together with the identity
\begin{equation}
(\Pbar+m)\gamma^{\mu}(\Pbar+m) =  2P^{\mu}(\Pbar+m)
\end{equation}
we obtain, for $P$ on-shell, 
\begin{eqnarray}
\lefteqn{\frac{1}{2E_\np}(\Pbar+m)
\frac{\partial\Sigma(P)}{\partial p_0}
(\Pbar+m)}
\nonumber\\
&=&
\frac{1}{E_\np}
\left[ m^2  \frac{\partial A(P)}{\partial p_0}
       E_\np^2\frac{\partial B(P)}{\partial p_0} 
      -\np^2\frac{\partial C(P)}{\partial p_0}
      +E_\np B_0(\np)
\right]_{p_0=E_\np} (\Pbar+m)
\nonumber\\
&=& \alpha(\np)(\Pbar+m) \ ,
\end{eqnarray}
where the definition of the function $\alpha(\np)$
in eq.~(\ref{alpha}) has been used.

Finally, collecting the above results, the derivative in 
eq.~(\ref{derivative}) is found to read
\begin{eqnarray}
\frac{d}{dp_0}
\left[\frac{1}{2p_0}(\Pbar+m)\Sigma(P)(\Pbar+m)\right]_{p_0=E_\np}
&=&
\frac{\Sigma_0(\np)}{E_\np}
\left[
\frac{\gamma_0 E_\np-m}{2E_\np}
(\Pbar+m)
+(\Pbar+m)
\frac{\gamma_0E_\np-m}{2E_\np}
\right]
\nonumber\\
&+&
\alpha(\np)
(\Pbar+m) \ .
\end{eqnarray}
Hence the following expression 
\begin{eqnarray}
\frac{d}{dp_0}
\left[
\frac{I^{\mu\nu}_{01}(H,P,Q)}{2p_0}\right]_{p_0=E_\np}
&=&
{\rm Tr}
\left\{
\Gamma^{\mu}(Q)
(\Hslash+m)
\Gamma^{\nu}(-Q)
\left[
\frac{\Sigma_0(\np)}{E_\np}
\frac{\gamma_0E_\np-m}{2E_\np}
+\frac{\alpha(\np)}{2}
\right]
(\Pbar+m)
\right\}
\nonumber\\
&+&
{\rm Tr}
\left\{
\Gamma^{\mu}(Q)
(\Hslash+m)
\Gamma^{\nu}(-Q)(\Pbar+m)
\left[
\frac{\Sigma_0(\np)}{E_\np}
\frac{\gamma_0E_\np-m}{2E_\np}
+\frac{\alpha(\np)}{2}
\right]
\right\}
\nonumber\\
\end{eqnarray}
yields the derivative of $I^{\mu\nu}_{01}(H,P,Q)$ with respect to $p_0$
and the similar result 
\begin{eqnarray}
\frac{d}{dh_0}
\left[
\frac{I^{\mu\nu}_{10}(H,P,Q)}{2h_0}\right]_{h_0=E_\nh}
&=&
{\rm Tr}
\left\{
\Gamma^{\mu}(Q)
\left[
\frac{\Sigma_0(\nh)}{E_\nh}
\frac{\gamma_0E_\nh-m}{2E_\nh}
+\frac{\alpha(\nh)}{2}
\right]
(\Hslash+m)
\Gamma^{\nu}(-Q)(\Pbar+m)
\right\}
\nonumber\\
&+&
{\rm Tr}
\left\{
\Gamma^{\mu}(Q)
(\Hslash+m)
\left[
\frac{\Sigma_0(\nh)}{E_\nh}
\frac{\gamma_0E_\nh-m}{2E_\nh}
+\frac{\alpha(\nh)}{2}
\right]
\Gamma^{\nu}(-Q)(\Pbar+m)
\right\} 
\nonumber\\
\end{eqnarray}
holds for the derivative of $I^{\mu\nu}_{10}(H,P,Q)$ with respect to $h_0$.
In addition, with the help of eq.~(\ref{spm}), we can write for 
on-shell momenta
\begin{eqnarray}
I^{\mu\nu}_{10}(H,P,Q)
&=&
{\rm Tr}
\left\{
\Gamma^{\mu}(Q)
(\Hslash+m)
\Sigma(H)
(\Hslash+m)
\Gamma^{\nu}(-Q)
(\Pbar+m)
\right\}
\nonumber\\
&=&
2m\Sigma_0(\nh)
{\rm Tr}
\left\{
\Gamma^{\mu}(Q)
(\Hslash+m)
\Gamma^{\nu}(-Q)
(\Pbar+m)
\right\}
\end{eqnarray}
and, as well, 
\begin{equation}
I^{\mu\nu}_{01}(H,P,Q)
=
2m\Sigma_0(\np)
{\rm Tr}
\left\{
\Gamma^{\mu}(Q)
(\Hslash+m)
\Gamma^{\nu}(-Q)
(\Pbar+m)
\right\} \ .
\end{equation}
Finally the response functions are found as
linear combinations of the diagonal components of the hadronic tensor, i.e. 
$W^{\mu\mu}$.
Using the above equations the latter reads
\begin{eqnarray}
W^{\mu\mu}
&=&
\int\frac{d^3h}{(2\pi)^3}
\frac{1}{4E_\nh E_\np}
{\rm Tr}
\left\{
\Gamma^{\mu}(Q)
\left[
\frac{\Sigma_0(\nh)}{E_\nh}
\frac{\gamma_0E_\nh-m}{2E_\nh}
+\frac{\alpha(\nh)}{2}
\right]
(\Hslash+m)
\Gamma^{\mu}(-Q)(\Pbar+m)
\right.
\nonumber\\
&&\kern 3cm
+\Gamma^{\mu}(Q)
(\Hslash+m)
\left[
\frac{\Sigma_0(\nh)}{E_\nh}
\frac{\gamma_0E_\nh-m}{2E_\nh}
+\frac{\alpha(\nh)}{2}
\right]
\Gamma^{\mu}(-Q)(\Pbar+m)
\nonumber\\
&&\kern 3cm
+\Gamma^{\mu}(Q)
(\Hslash+m)
\Gamma^{\mu}(-Q)
\left[
\frac{\Sigma_0(\np)}{E_\np}
\frac{\gamma_0E_\np-m}{2E_\np}
+\frac{\alpha(\np)}{2}
\right]
(\Pbar+m)
\nonumber\\
&&
\kern 3cm 
+\left.
\Gamma^{\mu}(Q)
(\Hslash+m)
\Gamma^{\mu}(-Q)(\Pbar+m)
\left[
\frac{\Sigma_0(\np)}{E_\np}
\frac{\gamma_0E_\np-m}{2E_\np}
+\frac{\alpha(\np)}{2}
\right]
\right\}
\nonumber\\
&&
\kern 3cm
\times
\delta(E_\nh+q_0-E_\np)
        \theta(k_F-h)
        \theta(p-k_F)
\nonumber\\
&+&
\int\frac{d^3h}{(2\pi)^3}
\frac{1}{4E_\nh E_\np}
{\rm Tr}
\left\{
\Gamma^{\mu}(Q)
(\Hslash+m)
\Gamma^{\mu}(-Q)
(\Pbar+m)
\right\}
\nonumber\\
&&
\times
\left(
\Sigma_0(\nh)
\frac{m}{E_\nh}
-\Sigma_0(\np)
\frac{m}{E_\np}
\right)
\frac{d}{d q_0}
\delta(E_\nh+q_0-E_\np)
        \theta(k_F-h)
        \theta(p-k_F) \ ,
\end{eqnarray}
which coincides with the result in eq.~(\ref{W-RSE}), obtained by
computing the response functions using the renormalized 
current and energy.

\section*{Appendix C. On-shell self-energy and field strength renormalization
function}

In this appendix we show in detail how to evaluate the on-shell 
self-energy in eq.~(\ref{Sigma_0}) and the field strength renormalization 
function in eq.~(\ref{alpha}). They can be expressed in terms of the integrals
$I(P)$ and $L^\mu (P)$ in eqs.~(\ref{I},\ref{Lmu}) as follows:
\be
\Sigma_0(\np)=2m B(E_\np,\np)
=-12m\frac{f^2}{m_\pi^2} \left[p_0 L_0(p_0,\np)-p L_3(p_0,\np)
-m^2 I(p_0,\np)\right]_{p_0=E_\np}
\label{Sigma0_app}
\ee
and
\begin{eqnarray}
\alpha(\np)
&=&
B_0(\np) + 
\frac{1}{E_\np}
\left[ 
       m^2\frac{\partial A(p_0,\np)}{\partial p_0}
     + E_\np^2\frac{\partial B(p_0,\np)}{\partial p_0}
     -\np^2\frac{\partial C(p_0,\np)}{\partial p_0}
\right]_{p_0=E_\np}
\nonumber\\
&=&
-12 m^2\frac{f^2}{m_\pi^2} \left[
\frac{L_0(p_0,\np)}{p_0}-I(p_0,\np)+\frac{\partial L_0(p_0,\np)}{\partial p_0}
-\frac{p}{p_0}\frac{\partial L_3(p_0,\np)}{\partial p_0}
-\frac{m^2}{p_0} \frac{\partial I(p_0,\np)}{\partial p_0}\right]_{p_0=E_\np} 
\ , \nonumber\\
\label{alpha_app}
\end{eqnarray}
where we have used eqs.~(\ref{A-definition}-\ref{C-definition}) and 
the derivatives
\begin{eqnarray}
\left(
\frac{\partial A(p_0,\np)}{\partial p_0}
\right)_{p_0=E_\np}
 &=&
-6\frac{f^2}{m_\pi^2}
\left[ 
          p_0 \frac{\partial L_0(p_0,\np)}{\partial p_0}
          -p \frac{\partial L_3(p_0,\np)}{\partial p_0}
          -m^2   \frac{\partial I(p_0,\np)}{\partial p_0} \right.
\nonumber \\
        & & \left.  + L_0 (p_0,\np)-p_0 I(p_0,\np)
\right]_{p_0=E_\np}
\\
\left(
\frac{\partial B(p_0,\np)}{\partial p_0}
\right)_{p_0=E_\np}
 &=&
-6\frac{f^2}{m_\pi^2}
\left[ 
          p_0 \frac{\partial L_0(p_0,\np)}{\partial p_0}
          -p \frac{\partial L_3(p_0,\np)}{\partial p_0}
          -m^2   \frac{\partial I(p_0,\np)}{\partial p_0}
\right]_{p_0=E_\np}
\\
\left(
\frac{\partial C(p_0,\np)}{\partial p_0}
\right)_{p_0=E_\np}
 &=&
-6\frac{f^2}{m_\pi^2}
\left[ 
          p_0 \frac{\partial L_0(p_0,\np)}{\partial p_0}
          -p \frac{\partial L_3(p_0,\np)}{\partial p_0}
          -m^2   \frac{\partial I(p_0,\np)}{\partial p_0} \right.
\nonumber \\
        & & \left.
          + L_0(p_0,\np)-\frac{p_0}{p}L_3(p_0,\np)
\right]_{p_0=E_\np} \, .
\end{eqnarray}

By choosing the z-axis in the
direction of $\np$ the angular integrals can be
performed analytically, yielding
\ba
I(E_\np,\np) &=& \frac{1}{(2\pi)^2}\int_0^{k_F} dk 
\frac{k}{4 p E_\nk}
    \ln\frac{\gamma(p,k)+2pk}{\gamma(p,k)-2pk}
\\
L_0(E_\np,\np)  &=& \frac{1}{(2\pi)^2}\int_0^{k_F} dk 
\frac{k}{4 p}
    \ln\frac{\gamma(p,k)+2pk}{\gamma(p,k)-2pk}
\\
L_3(E_\np,\np)  &=& \frac{1}{(2\pi)^2}\int_0^{k_F} dk 
\left\{
\frac{k^2}{2 p E_\nk}-\frac{k\gamma(p,k)}{8p^2E_\nk}
    \ln\frac{\gamma(p,k)+2pk}{\gamma(p,k)-2pk}
\right\}
\\
\left.\frac{\partial I(p_0,\np)}{\partial p_0}\right|_{p_0=E_\np}
 &=& \frac{1}{(2\pi)^2}\int_0^{k_F} dk
\frac{k}{2 p E_\nk} (E_\np-E_\nk) 
\left[\frac{1}{\gamma(p,k)+2pk}-\frac{1}{\gamma(p,k)-2pk}\right]
\\
\left.\frac{\partial L_0(p_0,\np)}{\partial p_0}\right|_{p_0=E_\np}
 &=& \frac{1}{(2\pi)^2}\int_0^{k_F} dk
\frac{k}{2 p} (E_\np-E_\nk) 
\left[\frac{1}{\gamma(p,k)+2pk}-\frac{1}{\gamma(p,k)-2pk}\right]
\\
\left.\frac{\partial L_3(p_0,\np)}{\partial p_0}\right|_{p_0=E_\np}
 &=& -\frac{1}{(2\pi)^2}\int_0^{k_F} dk
\frac{k}{4 p^2 E_\nk} (E_\np-E_\nk) 
\left\{
 \ln\frac{\gamma(p,k)+2pk}{\gamma(p,k)-2pk} \right.
\nonumber \\
& & \left.
+\gamma(p,k)\left[\frac{1}{\gamma(p,k)+2pk}-\frac{1}{\gamma(p,k)-2pk}\right]
\right\}\ ,
\ea
where we have defined the function
\be
\gamma(p,k) \equiv (E_\np-E_\nk)^2-
np^2-k^2-m_\pi^2=2m^2-m_\pi^2-2E_\np E_\nk\ .
\ee

By replacing the above integrals in eqs.~(\ref{Sigma0_app},\ref{alpha_app}) we obtain
\be
\Sigma_0(p) = \frac{3 m f^2}{2\pi^2 m_\pi^2} \int_0^{k_F} dk \frac{k^2}{E_\nk}
\left[
1+\frac{m_\pi^2}{4pk} \ln\frac{\gamma(p,k)+2pk}{\gamma(p,k)-2pk}
\right]
\ee
and
\be
\alpha(p) = \frac{3 m^2 f^2}{\pi^2 E_\np} \int_0^{k_F} dk \frac{k^2}{E_\nk}
\cdot\frac{E_\nk-E_\np}{\gamma^2(p,k)-4 p^2 k^2}\, .
\ee

It is interesting to note that for large $p$-values the following limits
hold
\ba
\lim_{p\to \infty}\alpha(p)&=& 0 \\
\lim_{p\to \infty}\Sigma_0(p) &=&
\frac{3 m f^2}{2\pi^2 m_\pi^2} \int_0^{k_F} dk \frac{k^2}{E_\nk}
\frac{3 m f^2}{4\pi^2 m_\pi^2}\left(E_Fk_F
-m^2\ln\frac{E_F+k_F}{m}\right) \, ,
\ea
where $E_F=\sqrt{k_F^2+m^2}$ is the Fermi energy. For $k_F=237$ MeV/c
the on-shell self-energy limit is $\sim 34$ MeV.


\end{document}